\pdfoutput=1

\documentclass[aps,twocolumn,showpacs,preprintnumbers,amsmath,amssymb,nofootinbib,superscriptaddress,showkeys,prd]{revtex4-1}

\usepackage{hyperref}
\usepackage{float}
\usepackage{epsfig}
\usepackage{graphicx}
\usepackage{xcolor}
\usepackage{latexsym}
\usepackage{natbib}
\usepackage{longtable}
\usepackage{dcolumn}
\usepackage{comment}

\usepackage{scrextend}

\usepackage[utf8]{inputenc}

\DeclareMathVersion{nxbold}
\SetSymbolFont{operators}{nxbold}{OT1}{cmr} {b}{n}
\SetSymbolFont{letters}  {nxbold}{OML}{cmm} {b}{it}
\SetSymbolFont{symbols}  {nxbold}{OMS}{cmsy}{b}{n}

\DeclareMathAlphabet{\mathcal}{OMS}{cmsy}{m}{n}
\usepackage{graphicx,color}
\usepackage{amsmath,amssymb,bm}
\usepackage{latexsym}
\usepackage{comment}
\usepackage{hyperref}
\usepackage{epsfig}

\begin{document}

\title{Partonic quasi-distributions of the proton and pion \\ from transverse-momentum distributions}

\author{Wojciech Broniowski}
\email{Wojciech.Broniowski@ifj.edu.pl}
\affiliation{The H. Niewodnicza\'nski Institute of Nuclear Physics, Polish Academy of Sciences, 31-342 Cracow, Poland}
\affiliation{Institute of Physics, Jan Kochanowski University, 25-406 Kielce, Poland}

\author{Enrique Ruiz Arriola}
\email{earriola@ugr.es}
\affiliation{Departamento de F\'isica At\'omica, Molecular y Nuclear and Instituto Carlos I de
  Fisica Te\'orica y Computacional,  Universidad de Granada, E-18071  Granada, Spain}
  
\begin{abstract}
The parton quasi-distribution functions
(QDFs) of Ji have been found by Radyushkin to be directly related to
the transverse momentum distributions (TMDs), to the pseudo-distributions,  and to the Ioffe-time
distributions (ITDs). This makes the QDF results at finite
longitudinal momentum of the hadron interesting in their own
right. Moreover, the QDF-TMD relation provides a gateway to the
pertinent QCD evolution, with respect to the resolution scale $Q$, for
the QDFs. Using the Kwieci\'nski evolution equations and well
established parameterizations at a low initial scale, we analyze the
QCD evolution of quark and gluon QDF components of the proton and the
pion. We discuss the resulting breaking of the longitudinal-transverse
factorization and show that it has little impact on QDFs at the
relatively low scales presently accessible on the lattice, but the
effect is visible in reduced ITDs at sufficiently large values of the
Ioffe time.  Sum rules involving derivatives of ITDs and moments of
the parton distribution functions (PDFs)
are {applied to the ETMC lattice data. This
allows us for a lattice determination of the transverse-momentum width of the
TMDs from QDF studies.}

\end{abstract}

\date{ver. 2, 5 February 2017}



\maketitle

\section{Introduction}

Partonic structure of hadrons is vividly exemplified experimentally
by the inclusive and semi-inclusive deep inelastic scattering, Drell-Yan
processes, the prompt-photon emission, etc., where abundant information
has been collected over the last 50 years. While parton distributions
are genuinely non-perturbative objects, the scaling violations, as dictated
by perturbative QCD (pQCD) radiative corrections  describing  the
{\it relative} scale dependence of the corresponding partonic
distributions, have been a major and lasting success of the theory at
sufficiently high resolution~\cite{Collins:2011zzd}. This verification
does not account for the {\it absolute} scale dependence of parton
distribution functions (PDFs), which are non-perturbative objects.  

Sound but isolated attempts have been undertaken on the
transverse lattice, formulated {\it directly} on the light
cone~\cite{Burkardt:2001mf,Dalley:2002nj} (for a review see,
e.g., \cite{Burkardt:2001jg}), which have incomprehensibly been abandoned or
forgotten. On the other hand, direct {\it ab initio} calculations involving
Euclidean lattices are precluded by the very Minkowski nature of
PDFs (the light-cone condition in the Minkowski space $t^2-z^2=0$ shrinks to one point,  $t_E^2+z^2=0$ in the Euclidean space where $t_E= i t$) 
and the unavoidable Lorentz symmetry breaking
of the finite lattice. Under those conditions, the only available
method for many years has been the computation of the lowest moments
of PDFs in the Bjorken $x$ variable.  
{Along this computational strategy, 
transverse momentum distributions 
(TMDs) on the lattice were  
pursued by Musch et al.~\cite{Musch:2010ka}  in a pioneering and comprehensive investigation. }

A more recent and promising breakthrough comes from an original
proposal by Ji~\cite{Ji:2013dva}, which provides an alternative route
to access PDFs directly from the Euclidean lattices and relies on the
so-called quasi-parton distribution functions (QDFs). These matrix
elements of partonic bilinears taken between hadron states moving at a
finite momentum $P_3$ were introduced as auxiliary objects. They
involve boosting space-like correlators to a finite momentum and,
eventually, may be used to extrapolate the results to the
infinite-momentum frame, $P_3 \to \infty$, yielding PDFs. Many
theoretical
discussions~\cite{Xiong:2013bka,Ji:2014gla,Ma:2014jla,Ji:2015jwa,Ji:2015qla,Radyushkin:2016hsy,Monahan:2016bvm,Chen:2016fxx,%
Ji:2017rah,Chen:2017mzz,Carlson:2017gpk,Briceno:2017cpo,Rossi:2017muf,Stewart:2017tvs,Radyushkin:2017ffo},
lattice
simulations~\cite{Alexandrou:2015rja,Alexandrou:2016eyt,Alexandrou:2016jqi,Alexandrou:2017dzj,Orginos:2017kos}
and quark-diquark model calculations~\cite{Gamberg:2014zwa} have been
undertaken along these lines.

Quite generally, the full partonic structure contains both the
longitudinal and transverse information, which can equivalently be
described in terms of different kinematic variables. Fourier
transformations generate a proliferation of possible definitions of
these objects, depending on the chosen variables, whereas relativistic
covariance provides relations between them (for instance, {\em
transversity relations}, connecting the Light-Cone (LC) and Equal-Time
(ET) wave functions of the pion~\cite{Miller:2009fc,Broniowski:2009dt,Arriola:2010up,Miller:2010nz}).

In a series of remarkable and insightful papers,
Radyushkin~\cite{Radyushkin:2016hsy,Radyushkin:2017cyf,Radyushkin:2017gjd,Radyushkin:2017lvu}
unveiled a fundamental connection between Ji's QDFs and the
well-studied TMDs~\cite{Collins:2011zzd} (see,
e.g.,~\cite{Angeles-Martinez:2015sea} for an overview) and the
honorable Ioffe-time-distributions
(ITDs)~\cite{Ioffe:1969kf,Braun:1994jq}.  The relation follows just
from the Lorentz covariance (and from projecting out the subleading
twist structures).  This observation has triggered incipient further
works on the
lattice~\cite{Orginos:2017kos,Karpie:2017bzm,Monahan:2017oof} {
providing in addition a different and upgraded perspective to former
TMD lattice studies~\cite{Musch:2010ka}}. These crucial findings show
that QDFs are in fact complementary to TMDs, thus QDFs, even at low
values of $P_3$, should not be viewed as mere auxiliary mathematical
devices, but rather as physical objects interesting in their own
right.  The wealth of information on TMDs from phenomenological
studies in the so-called $k_T$-factorization scheme is therefore
inherited by QDFs.  Besides, this connection provides a handle on the
issue of the resolution scale dependence, since much is already known
on TMDs from the pQCD evolution aspect. Moreover, the results for QDFs
at finite $P_3$ are interesting for testing non-perturbative models of
the proton and pion structure.

Within the standard folklore of the TMD phenomenological studies, the
independence of the longitudinal and transverse dynamics has been
implemented through a Gaussian factorization ansatz, which {\em a
fortiori} complies to the Drell-Yan~\cite{DAlesio:2004eso} and
semi-inclusive deep-inelastic scattering
investigations~\cite{Schweitzer:2010tt}, as well as to the recent lattice
studies~\cite{Musch:2010ka}. This important issue has recently been reanalyzed
and confirmed for the ITDs on the quenched lattice~\cite{Orginos:2017kos,Karpie:2017bzm}.

The purpose of this paper is to discuss certain aspects of the QDF-TMD
connection which are potentially relevant for phenomenological and
lattice studies, but have not yet been covered to sufficient detail in
the literature. A careful scrutiny of the longitudinal-transverse
factorization is one of the key issues we present here. Thanks to the
Radyushkin QDF-TMD relation, one may investigate the QCD evolution of
QDFs with a probing scale $Q$  via the known methods of the TMD
evolution.\footnote{The correct definition of a
parton density requires a specification of the resolution scale, which
will generically be denoted by $Q$.}  Specifically, we use here a simple scheme based on  the
Ciafaloni, Catani, Fiorani, and Marchesini (CCFM)
framework~\cite{Ciafaloni:1987ur,Catani:1989yc,Catani:1989sg},
developed long ago for the then so-called $k_T$-{\em unintegrated}
gluon distributions to evolve TMDs.  The CCFM equations in the single
loop approximation were later adapted to include quarks by
Kwieci\'nski~\cite{Kwiecinski:2002bx} (see
also~\cite{Gawron:2002kc,Gawron:2003qg,RuizArriola:2004ui}).  We use
the solutions of the Kwieci\'nski evolution equations for both the
proton and the pion, where the initial condition for the evolution
imposed at the scale $Q_0$ is obtained by assuming a factorized ansatz
involving a known parametrization of the PDFs and a choice of the
transverse-coordinate profile function.  We bring up the fact that the
QCD evolution of TMDs precludes factorization at {\em all}
scales. However, the induced breaking does not generate a large effect
on the QDFs at the relatively low values of $Q \sim 2$~GeV, which are
presently available on the lattice.

The factorization breaking from the QCD evolution  
is visible in ITDs at magnitudes of the Ioffe time above several units, thus in the tail, 
which via Fourier transform corresponds to low values of $x$.
Therefore, the factorization breaking becomes relevant at low values of $x$ and 
is enhanced at higher values of $Q$. Note, however, that the low-$x$
domain is not accessible to the methodology of the present Euclidean lattice investigations.
We also explore the reduced ITDs proposed in~\cite{Orginos:2017kos}, which are specifically 
designed to probe the longitudinal-transverse factorization. With the factorization breaking induced by the 
Kwieci\'nski evolution, we find effects in the tails of the reduced ITDs, which become increasingly relevant as the 
value of the longitudinal momentum of the hadron is reduced.

In our study, we provide QDFs for both quarks and gluons in the proton
and the pion, as well as the corresponding ITDs. One should keep in
mind, however, that an evaluation of the gluon distributions on the
lattice is more demanding than for the quark case.

On the general ground, we spell out simple sum rules linking the
derivatives of ITDs at the origin to the $x$-moments of the PDFs and
the moments of the $k_T$ distribution. These sum rules may be useful
for consistency checks of the lattice results. For the reduced ITDs,
they set the slope of the imaginary part and the curvature of the real
part at the origin, which are universal, and determined by the first
and second $x$-moment of the corresponding PDF.  They also link
in a simple way the $x$ moments of the QDFs and PDFs,
and the $k_T$ moments of TMDs.  We have applied the sum rules to the
lattice data of~\cite{Alexandrou:2016eyt}, confirming proper scaling
with $P_3$ and extracting the with of the $k_T$ distribution.

\section{Definitions and relations \label{sec:def}}

We begin by presenting a glossary of relevant definitions and
formulas. The results referring to the Ioffe distributions and the
link between QDFs and TMDs were obtained in
previous works~\cite{Braun:1994jq,Radyushkin:2016hsy,Radyushkin:2017gjd,Orginos:2017kos}.
We review them here for completeness and to establish our notation.

\subsection{Quark distributions}

The Lorentz covariance allows one to parametrize the matrix elements of the spin-averaged quark bilinears as
\begin{eqnarray}
&&\hspace{-4mm} \langle P | \bar \psi(0) \gamma^\mu  U[0,z] \psi(z) | P \rangle \nonumber \\
&&=  P^\mu h( P\cdot z, z^2) + z^\mu h_z( P\cdot z, z^2), \label{eq:fdef}
\end{eqnarray}
where $| P \rangle $ is a hadron state of four-momentum $P$, the
link operator, providing the gauge invariance, is denoted as $U[0,z]$,
and $h( P\cdot z, z^2)$ and $h_z( P\cdot z, z^2)$ are scalar
functions. The term proportional to $z^\mu$ in the decomposition
of Eq.~(\ref{eq:fdef}) contains subleading twist pieces only, so it is favorable
to project it out from the following definitions~\cite{Radyushkin:2016hsy,Radyushkin:2017gjd}. 
The issue is discussed in some greater detail in Appendix~\ref{app:decomp}.

Following~\cite{Radyushkin:2016hsy,Radyushkin:2017gjd,Orginos:2017kos}, we define the parton
quasi-distributions (QDFs) analogously to the original proposal by
Ji~\cite{Ji:2013dva}, but retaining the $P^\mu$ term only, i.e.,
\begin{eqnarray}
\tilde q(y,P_3) =  P_3 \int \frac{dz_3}{2\pi} e^{-i y P_3 z_3}  h(-P_3 z_3,-z_3^2). \label{eq:qh}
\end{eqnarray}
Here $y$ acquires the interpretation of the fraction of the hadron's longitudinal
momentum $P_3$ carried by the parton, with the support $y \in
(-\infty,\infty)$.  As shown by Ji~\cite{Ji:2013dva}, in the limit of
$P_3 \to \infty$ one recovers the usual PDFs,
\begin{eqnarray}
 \lim_{P_3 \to \infty} \tilde q(y,P_3) = q(x=y), \label{eq:limit}
\end{eqnarray}
where
\begin{eqnarray}
q(x) =  P_+ \int \frac{dz_-}{2\pi} e^{i x P_+ z_-}  h(P_+ z_-,0),
\end{eqnarray}
with $y=x$ denoting the fraction of the light-front momentum of the hadron carried by the parton. 

More precisely, in the adopted convention the distribution for $x \in
[0,1]$ corresponds to the quarks, and for $x \in [-1,0]$ to the
anti-quarks, i.e., $\bar q(x) = - q(-x)$~\cite{Jaffe:1983hp} (see
Ref.~\cite{Jaffe:1985je} for a pedagogical introduction). Then, for
the valence and sea quarks one has
\begin{eqnarray}
&& q_{\rm val}(x)=q(x)-\bar q(x) = q(x)+q(-x), \;\;\;\ x \in [0,1], \nonumber \\
&& q_{\rm sea}(x)= \left \{ \begin{array}{cl}\bar q(x) = -q(-x)  &  
\;\;\;\ {\rm for~~}x \in [0,1], \\ -\bar q(-x)=  q(x) & \;\;\;\ {\rm for~~}x \in [-1,0].   \end{array} \right . \label{eq:dom}
\end{eqnarray}

The transverse-momentum unintegrated parton distribution, or TMD, is
defined as
\begin{eqnarray}
q(x,k_1,k_2) &\equiv& P^+ \!\!\! \int \frac{dz^-}{2\pi}e^{i x P_+ z_-}
\!\!\!\int \frac{dz_1}{2\pi} e^{i k_1 z_1}
\!\!\!\int \frac{dz_2}{2\pi} e^{i k_2 z_2} \nonumber \\
&\times & h(P_+ z_-,-z_1^2-z_2^2).  \label{eq:TMD}
\end{eqnarray}
From the axial symmetry $q(x,k_1,k_2)=q(x,k_T^2)$, with $\vec{k}_T = (k_1,k_2)$.

\subsection{Gluon distributions}

For the gluons, the corresponding matrix element can be defined analogously as 
\begin{eqnarray}
&& \hspace{-8mm} \langle P | F^{\mu \alpha}(0)  U[0,z] {F_{\alpha}}^\nu(z) | P \rangle z_\mu z_\nu \nonumber \\ && = P^\mu P^\nu h_g( P\cdot z, z^2) + \dots ,
\label{eq:fdefG}
\end{eqnarray}
with the dots denoting terms containing higher twists only, 
and the QDF and PDF, multiplied by the corresponding momentum fractions, are defined as
\begin{eqnarray}
&& y \tilde g (y,P_3) =  P_3 \int \frac{dz_3}{2\pi} e^{-i y P_3 z_3}  h_g(-P_3 z_3,-z_3^2), \nonumber \\
&& x g(x) =  P_+ \int \frac{dz_-}{4\pi} e^{i x P_+ z_-}  h_g(P_+ z_-,0).
\label{eq:qhG}
\end{eqnarray}
The quasi-distribution $y \tilde g(y,P_3)$ is distributed symmetrically in
$y \in (-\infty,\infty)$, whereas $x g(x)$ is distributed
symmetrically in the domain $x \in [-1,1]$ (see, e.g.,
Refs.~\cite{Ji:1998pc} for discussion).  Then, together with the quark
and antiquark distributions they form the singlet component of the
partonic distributions in context of their QCD evolution.

\subsection{Transversity relations}

Lorentz invariance of the matrix elements allows one to obtain
relations, which otherwise are a priori not obvious. To our knowledge,
the first investigations along these lines were done
in~\cite{Miller:2009fc,Broniowski:2009dt,Arriola:2010up,Miller:2010nz} for the case of the pion
wave function (see Appendix~\ref{sec:tr} for a brief review).
The functions $\Phi_a( \alpha, z^2 )$ of Eq.~(\ref{eq:Phi}) are analogs of the {\em pseudo-distributions} introduced by
Radyushkin~\cite{Radyushkin:2017cyf} and advocated as a basic entity of the formalism.

Note that the functional dependence in both integrands appearing in
the QDF in Eq.~(\ref{eq:qh}) and the TMD in Eq.~(\ref{eq:TMD}) suggests
a direct link. Radyushkin~\cite{Radyushkin:2016hsy} showed that 
QDFs are simply but non-trivially related to TMDs,
\begin{eqnarray} 
\tilde q(y,P_3)= P_3 \int dk_1 \!\!\int dx \,q(x,k_1^2+(x-y)^2P_3^2). \label{eq:rad}
\end{eqnarray}
For completeness, in Appendix~\ref{sec:rad} we review the derivation
of the Radyushkin relation from the Lorentz
invariance~\cite{Radyushkin:2017cyf} in an explicit manner.

We may use the transverse coordinate representation (Fourier-conjugate to definition~(\ref{eq:TMD}) and 
denoted with a hat) of the TMD, 
\begin{eqnarray}
\hspace{-4mm} \hat q(x,z_T^2) = \!\!\! \int \frac{d\nu}{2\pi}e^{-i \nu x}  h(-\nu,-z_T^2),  \label{eq:TMDz0}
\end{eqnarray}
where the transverse coordinate is $z_T=(0,z_2)$, whereas the integration variable $\nu=- P\cdot z$ is the Ioffe time~\cite{Ioffe:1969kf,Braun:1994jq}.
In the Lorentz-invariant notation one recovers Radyushkin's pseudo-distribution ${\cal P}$~\cite{Radyushkin:2017cyf}
\begin{eqnarray}
\hspace{-4mm} \hat q(x,-z^2) \equiv {\cal P}(x,-z^2)= \!\!\! \int \frac{d\nu}{2\pi}e^{-i \nu x}  h(-\nu,z^2),  \label{eq:TMDz0L}
\end{eqnarray}
which in the frame $z=(0,0,0,z_3$) applied below becomes
\begin{eqnarray}
\hspace{-4mm} \hat q(x,z_3^2) = {\cal P}(x,z_3^2)= \!\!\! \int \frac{d\nu}{2\pi}e^{-i \nu x}  h(-\nu,-z_3^2).  \label{eq:TMDz}
\end{eqnarray}
We can now write down an equivalent form of Eq.~(\ref{eq:rad}), which links QDF to TMD or to the pseudo-distribution, namely
\begin{eqnarray} 
\hspace{-7mm}  \tilde q(y,P_3)&=& P_3 \int dx \int \frac{dz_2}{2\pi} e^{-i(y-x)z_2 P_3}\hat q(x,z_T^2)  \nonumber \\
&=& P_3 \int dx \int \frac{dz_3}{2\pi} e^{-i(y-x)z_2 P_3}\hat q(x,z_3^2).
\label{eq:b}
\end{eqnarray}
These relations can be inverted if we invoke the integration over $P_3$:
\begin{eqnarray} 
\hspace{-7mm} &&  \hat q(x,z_T^2) = z_2 \int dy\,   \int dP_3 e^{i(y-x)z_2 P_3} \tilde q(y,P_3), \nonumber \\
 &&  \hat q(x,z_3^2) = z_3 \int dy\,   \int dP_3 e^{i(y-x)z_3 P_3} \tilde q(y,P_3).
\label{eq:inv}
\end{eqnarray}
Therefore the knowledge of quasi-distributions at all values of the
hadron momentum $P_3$ allows one for obtaining the corresponding TMD and
the pseudo-distribution in $z_3$.\footnote{As
remarked in~\cite{Orginos:2017kos}, the implicit prescription for the
Wilson gauge link is a straight line extending from $0$ to $z_3$,
rather than the semi-infinite stapled-link form~\cite{Boer:2011xd}.
Similar prescriptions are used in the lattice studies of
TMDs~\cite{Musch:2010ka} or
QDFs~\cite{Alexandrou:2015rja,Alexandrou:2016eyt,Alexandrou:2016jqi,Alexandrou:2017dzj,Orginos:2017kos}.}

The matrix element $h(-\nu, z^2)$ appearing in Eq.~(\ref{eq:TMDz}) is referred to as the Ioffe-time distribution (ITD)~\cite{Braun:1994jq,Orginos:2017kos}, 
and is equal to $2{\cal M}_p(\nu,-z^2)$ in the notation of~\cite{Radyushkin:2016hsy,Radyushkin:2017gjd}.
The {\em normalized amplitude}~\cite{Musch:2010ka}, or the {\em reduced}\, ITD~\cite{Orginos:2017kos}, 
used to probe the transverse-longitudinal factorization, is defined as
\begin{eqnarray}
\mathfrak{M}(\nu,-z^2)=\frac{{\cal M}_p(\nu,-z^2)}{{\cal M}_p(0,-z^2)}=\frac{h(-\nu,z^2)}{h(0,z^2)}. \label{eq:red}
\end{eqnarray}
The denominator has an interpretation of the rest-frame distribution. 

This definition has the advantage that the self-energy of the Wilson loop characterized by a
multiplicative renormalization factor $\sim e^{-z_3 m }$ cancels in
the ratio. This finding on the lattice~\cite{Orginos:2017kos} is in
harmony with the improved parton quasi-distribution through the Wilson
line renormalization~\cite{Chen:2016fxx}, which safely removes power
divergences ubiquitous in lattice QCD. 

\section{Sum rules for the matrix elements of bilocal fields \label{sec:sumrules}} 

Fourier inversion  with $\nu=- P_3 z_3$  of Eq.~(\ref{eq:TMDz})  yields
the relation of the ITD with the pseudo-distribution,
\begin{eqnarray}
h(-P_3 z_3,-z_3^2) =  \int_{-1}^1 {dx}\,e^{i P_3 z_3  x} \hat q(x,z_3^2),
\label{eq:TMDzinv1}
\end{eqnarray}
whereas the corresponding inversion of Eq.~(\ref{eq:qh}) links ITD to QDF,
\begin{eqnarray}
h(-P_3 z_3,-z_3^2) =  \int_{-\infty}^\infty {dy}\,e^{i P_3 z_3  y} \tilde q(y,P_3).
\label{eq:TMDzinv2}
\end{eqnarray}
We immediately see that the real part of $h$ is an even function of $z_3$, whereas the imaginary part is odd.  
Note that according to Eq.~(\ref{eq:dom}), the valence quarks contribute both to the real and imaginary parts of $h$, 
the sea quarks contribute to the imaginary part of $h$ only, and the gluons yield $h_g$ which is real.
Also, Eqs.~(\ref{eq:TMDzinv1}) and (\ref{eq:TMDzinv2}) immediately yield the equality 
\begin{eqnarray}
\hspace{-7mm} \int_{-1}^1 {dx}\,e^{i P_3 z_3  x} \hat q(x,z_3^2) = \int_{-\infty}^\infty {dy}\,e^{i P_3 z_3  y} \tilde q(y,P_3),
\label{eq:sr0}
\end{eqnarray}
which leads to the new sum rules presented shortly.

The normalization condition for the quark PDF yields, from Eqs.~(\ref{eq:TMDzinv1},\ref{eq:TMDzinv2}),
\begin{eqnarray}
&& \hspace{-4mm}  h(0,0)  =  \int_{-1}^1\!\! {dx}\, \hat q(x,0) =  \int_{-1}^1\!\! {dx} \, q(x) \nonumber \\
&& =  \int_{-\infty}^\infty\!\! {dy} \, \tilde q(y,P_3)= N_q,  \label{eq:snorm}
\end{eqnarray}
where $N_q$ is the number of valence quarks of a given flavor. 

Taking subsequent derivatives of the left- and right-hand sides of Eq.~(\ref{eq:sr0}) 
with respect to $z_3$ at the origin, under the assumption
of regularity of $\hat q(x,z_3^2)$ in $z_3^2$, 
yields simple sum rules which depend parametrically on $P_3$. 

The first derivative of Eq.~(\ref{eq:sr0}) is related to fractions of momenta carried by the quarks, 
\begin{eqnarray}
 && \left . \frac{d}{dz_3} h(-P_3 z_3,-z_3^2) \right |_{z_3=0} \nonumber \\ 
 && = i P_3  \int_{-1}^1 \!\!  {dx}\, x \, q(x) =  i P_3  \int_{-\infty}^\infty \!\!  {dy}\, y \, \tilde q(y,P_3)  \label{eq:deriv0}
\end{eqnarray}
(we have used the fact that $\left . {d} \hat q(x,z_3^2)/{dz_3} \right |_{z_3=0}=0$, which follows from regularity),
or
\begin{eqnarray}
\langle x \rangle_q  = \langle y \rangle_q(P_3) = \langle y \rangle_q \label{eq:deriv}
\end{eqnarray}
(the brackets denote the moments appearing in Eq.~(\ref{eq:deriv0})).

We notice from Eq.~(\ref{eq:deriv0}) that the derivative of the imaginary part of $h$ with respect to $z_3$ at the origin is proportional to $P_3$
and contains a known coefficient, $\langle x \rangle_q$.
We also note that the first $y$-moment of $\tilde q(y,P_3)$, which in principle might depend on $P_3$, in fact does not, as 
indicated in Eq.~(\ref{eq:deriv}).

Similarly, the second derivative of Eq.~(\ref{eq:sr0}) 
with respect to $z_3$ at the origin yields  
\begin{eqnarray}
&&  \left . \frac{d^2}{dz_3^2} h(-P_3 z_3,-z_3^2) \right |_{z_3=0} \label{eq:deriv20} \\
&& = - P_3^2  \int_{-1}^1 \!\!  {dx}\, x^2 q(x) + \int_{-1}^1 \!\!  {dx}\,  \left . \frac{d^2}{dz_3^2} \hat q(x,z_3^2) \right |_{z_3=0} \nonumber \\
&& =  - P_3^2  \int_{-\infty}^\infty \!\!  {dy}\, y^2 \tilde q(y,P_3). \nonumber
\end{eqnarray}
Since the quasi-distributions and the TMDs have the same functional form, their Maclaurin expansions, correspondingly, 
in $z_3$ or $z_2$ are the same. The interpretation of the coefficients can thus be given via the ($x$-dependent)  $k_T$-moments of the TMDs. 
In particular, for the quadratic term we have
\begin{eqnarray}
&&  \left . \frac{d^2}{dz_3^2} \hat q(x,z_3^2) \right |_{z_3=0}= \left . \frac{d^2}{dz_2^2} \hat q(x,z_2^2) \right |_{z_2=0} \nonumber \\
&& =- \frac{1}{2} P_3^2 \langle k_T^2 \rangle(x) \, q(x). \label{eq:kTdef}
\end{eqnarray}
We introduce the short-hand notation for the $x$-averaged $k_T$ width per valence quark,  
\begin{eqnarray}
\overline{\langle k_T^2 \rangle} = \int_{-1}^1 \! {dx} \langle k_T^2\rangle (x)  q(x)/N_q.
\end{eqnarray}
We may now rewrite Eq.~(\ref{eq:deriv20}) in a compact form
\begin{eqnarray}
\hspace{-7mm}  \langle x^2 \rangle_q +\frac{N_q  \overline{\langle k_T^2 \rangle}}{2P_3^2}  
 =  \langle y^2 \rangle_q(P_3).  \label{eq:deriv2}
\end{eqnarray}
We note from Eq.~(\ref{eq:deriv20}) that increasing $P_3$ makes the function $h$ more and more sharply peaked at the origin. 
Also, the width of QDF is larger than the width 
of the corresponding PDF, as follows from 
the relation of the second moments~(\ref{eq:deriv2}), with the first moments being equal, cf.~Eq.~(\ref{eq:deriv}).
The effect is of the order $\overline{\langle k_T^2 \rangle} /P_3^2$, 

Higher-order relations may be readily obtained taking more
differentiations with respect to $z_3$, and hold as long 
as the obtained moments exist.

For the gluon distributions, analogously, 
\begin{eqnarray}
&& \hspace{-4mm} h_g(0,0) = \langle x \rangle_g = \langle y \rangle_g, \nonumber \\ 
&& \hspace{-4mm} \left . \frac{d}{dz_3} h_g(-P_3 z_3,-z_3^2) \right |_{z_3=0}=0,  \label{eq:derivG}
\end{eqnarray}
and
\begin{eqnarray}
&&  \left . \frac{d^2}{dz_3^2} h_g(-P_3 z_3,-z_3^2) \right |_{z_3=0} \label{eq:deriv2G} \\
&& = - P_3^2 \langle x^3 \rangle_g -  \frac{1}{2} \int_{-1}^1 \!\!  {dx} \langle k_T^2(x) \rangle_g x g(x)  = - P_3^2 \langle y^3 \rangle_g(P_3).  \nonumber
\end{eqnarray}

Equations (\ref{eq:deriv0}-\ref{eq:deriv2G}) may have a practical significance in the interpretation and consistency checks 
of the lattice data. The consistency can be verified by 
checking the $P_3$ dependence in Eq.~(\ref{eq:deriv}) with the known 
$x$-moment. 
Equations~(\ref{eq:deriv2},\ref{eq:deriv2G}) provide
a way to effectively measure the average spreading of the transverse momentum in the TMDs. One would 
need to obtain the matrix elements $h$ or $h_g$ at various values of $P_3$ with a
sufficient accuracy, such that interpolation fits can be made and then derivatives at the origin taken. 
In Sec.~\ref{sec:complat} we successfully apply the sum rules to the lattice data from~\cite{Alexandrou:2016eyt}. 

The distributions in the Ioffe time $\nu=-P_3 z_3$ display more universality, as then the slope of the imaginary part of $h$ at the origin
is common to all values of $P_3$,
\begin{eqnarray}
\hspace{-4mm} \left . \frac{d}{d\nu} h\left(-\nu,-\frac{\nu^2}{P_3^2}\right)  \right |_{\nu=0} = i \langle x \rangle_q = i \langle y \rangle_q,  \label{eq:derivnu}
\end{eqnarray}
whereas the curvature at the origin of the real part of $h$ is
\begin{eqnarray}
&& \left . \frac{d^2}{d\nu^2} h\left(-\nu,-\frac{\nu^2}{P_3^2}\right) \right |_{\nu=0} \!\!\!\!\!\! 
= -  \langle x^2 \rangle_q - \frac{N_q \overline{\langle k_T^2 \rangle}}{2P_3^2} = -  \langle y^2 \rangle_q(P_3). \nonumber \\ \label{eq:deriv2bnu}
\end{eqnarray}
Above, we have used the same method as in the derivation of Eq.~(\ref{eq:deriv},\ref{eq:deriv2}).

There is even more vivid universality for the reduced ITDs, where both first and second derivatives at the origin are independent of $P_3$:
\begin{eqnarray}
\hspace{-4mm} && \left . \frac{d}{d\nu}\mathfrak{M}(\nu,\nu^2/P_3^2) \right |_{\nu=0} = i \langle x \rangle_q = i \langle y \rangle_q, \label{eq:dersc} \\
&& \left . \frac{d^2}{d\nu^2} \mathfrak{M}(\nu,\nu^2/P_3^2) \right |_{\nu=0} \!\!\!\!\!\! 
= -  \langle x^2 \rangle_q   = -  \langle y^2 \rangle_q(P_3)+ \frac{N_q \overline{\langle k_T^2 \rangle}}{2P_3^2}. \nonumber
\end{eqnarray}

The discussed universality behavior was observed in actual (quenched)
lattice simulations reported in~\cite{Orginos:2017kos}.

\section{Factorization ansatz \label{sec:factor}} 

In modeling of TMDs, a popular assumption is the factorization ansatz
\begin{eqnarray}
q(x,k_T) = q(x) F(k_T), 
\end{eqnarray}
or, equivalently,
\begin{eqnarray}
\hat q(x,z_T) = q(x) \hat F(z_T), \label{eq:fact}
\end{eqnarray}
which separates the transverse and longitudinal dynamics (we will
discuss later on the departures from this assumption). Whereas this
has traditionally been an out-of-ignorance guess, lattice calculations
of TMDs speak in favor of this factorization, at least as long as
\mbox{$m_\pi \simeq 600~{\rm MeV}$}~\cite{Musch:2010ka}.\footnote{Quite
surprisingly, this {\it a priori} naive property is indeed violated for
the spectator~\cite{Bacchetta:2008af} and chiral quark soliton
models~\cite{Wakamatsu:2009fn} for the proton, as well as for the
chiral quark models for the
pion~\cite{Weigel:1999pc,RuizArriola:2003wi,RuizArriola:2003bs,Noguera:2015iia} away from the strict chiral limit.}
Moreover, one typically  uses a Gaussian shape 
\begin{eqnarray}
\hspace{-5mm} 
F(k_T)=\frac{e^{-\frac{k_T^2}{\langle k_T^2 \rangle}}}{\pi \langle k_T^2 \rangle}, \;\;  \hat F(z_T)=e^{-\frac{z_T^2}{2\sigma_0^2}},
\;\; \sigma_0^2 =\frac{2}{\langle k_T^2 \rangle}. \label{eq:gauss}
\end{eqnarray}

The Gaussian factorization ansatz has been favorably checked against the
data in the Drell-Yan~\cite{DAlesio:2004eso} and semi-inclusive
deep-inelastic scattering~\cite{Schweitzer:2010tt}. In the context of
quasi-distributions, this form was explored in~\cite{Radyushkin:2016hsy,Radyushkin:2017gjd,Orginos:2017kos}. 
A typical value of $\langle k_T^2 \rangle$ extracted from
phenomenological studies (at energy scales of a few GeV) is
$\langle k_T^2 \rangle \sim 0.3-0.6~{\rm GeV}^2$~\cite{Melis:2014pna,Bacchetta:2017gcc}.

\begin{figure}[tb]
\begin{center}
\includegraphics[angle=0,width=0.45 \textwidth]{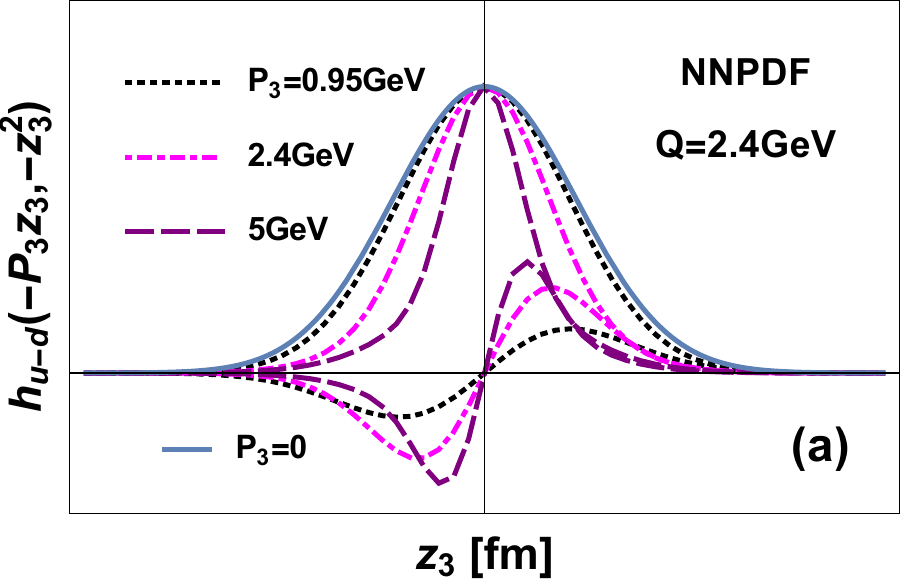}\\
\vspace{3mm}
\includegraphics[angle=0,width=0.45 \textwidth]{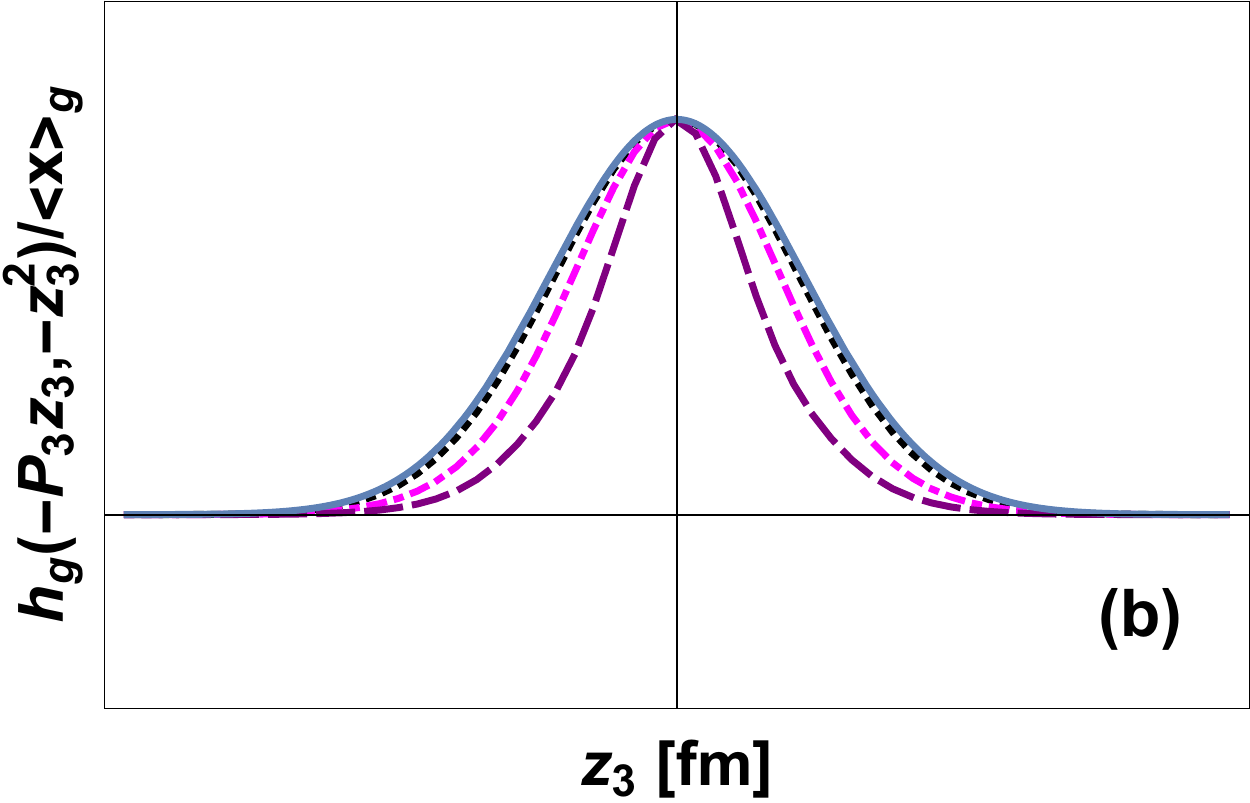}
\end{center}
\vspace{-5mm}
\caption{Matrix element corresponding to the (a) quark (b) gluon QDF of the proton for several values of $P_3$, 
evaluated in the factorization model, where 
the PDFs at the scale $Q=2.4$~GeV, taken from the NNPDF parametrization, are supplied with a Gaussian form factor 
with the width parameter $\langle k_T^2 \rangle=0.6$~GeV.
The real parts are symmetric in $z_3$, whereas the imaginary parts are 
antisymmetric. The solid line ($P_3=0$) indicates the form factor $\hat F(z_3^2)$.  \label{fig:Ioffe1}}
\end{figure}

\begin{figure}[tb]
\begin{center}
\includegraphics[angle=0,width=0.45 \textwidth]{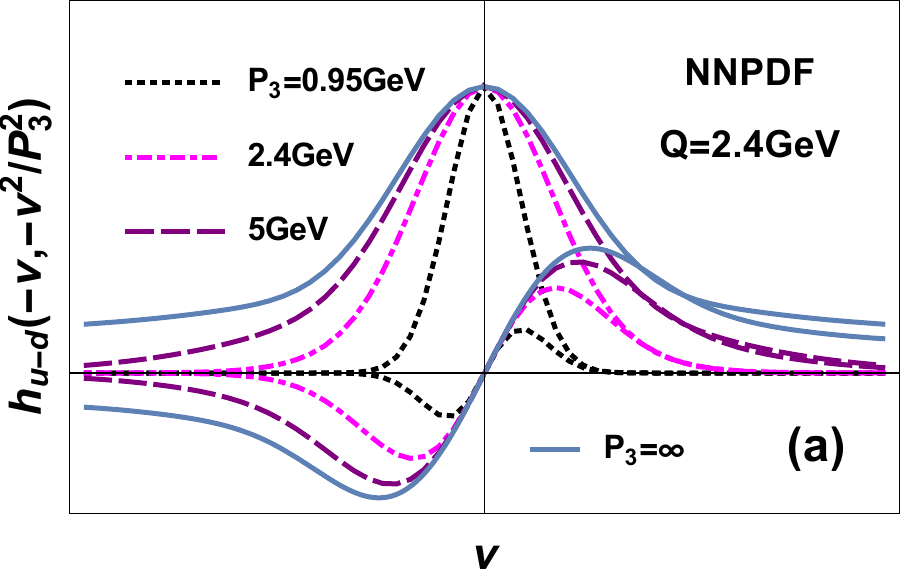}\\
\vspace{3mm}
\includegraphics[angle=0,width=0.45 \textwidth]{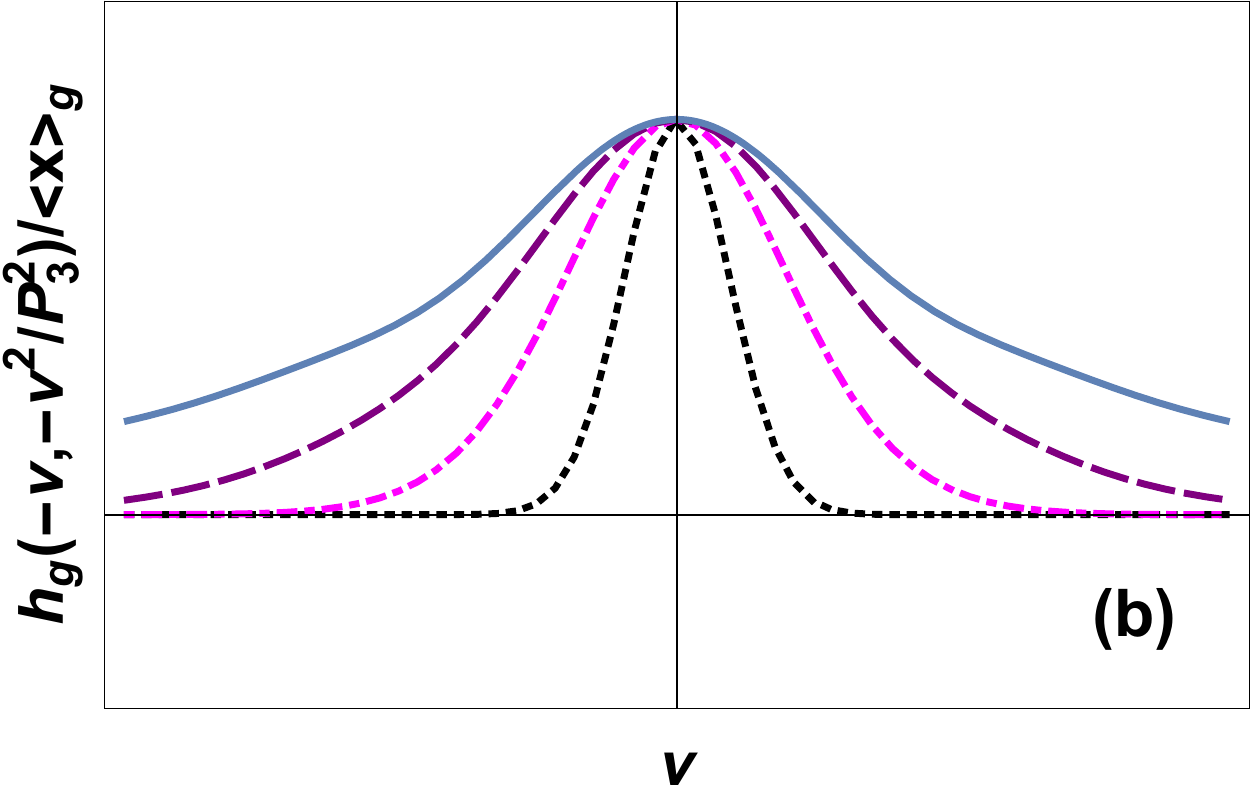}
\end{center}
\vspace{-5mm}
\caption{Same as in Fig.~\ref{fig:Ioffe1} but for the the distributions in the Ioffe time $\nu =P_3 z_3$.
In this case the solid lines labeled $P_3=\infty$ represent the limits of Eq.~(\ref{eq:ITDi}). \label{fig:ioffe2}}
\end{figure}

\begin{figure}[tb]
\begin{center}
\includegraphics[angle=0,width=0.45 \textwidth]{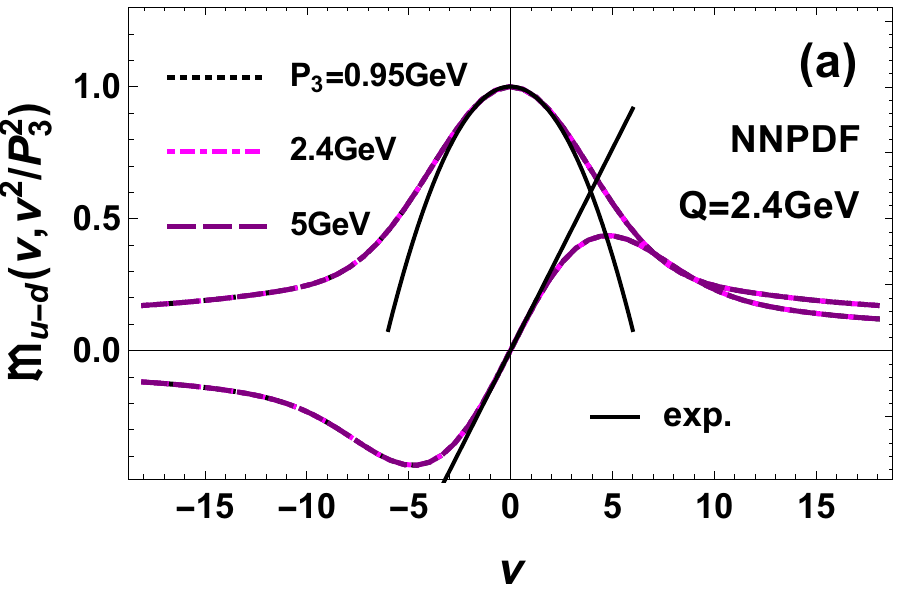} \\
\vspace{3mm}
\includegraphics[angle=0,width=0.45 \textwidth]{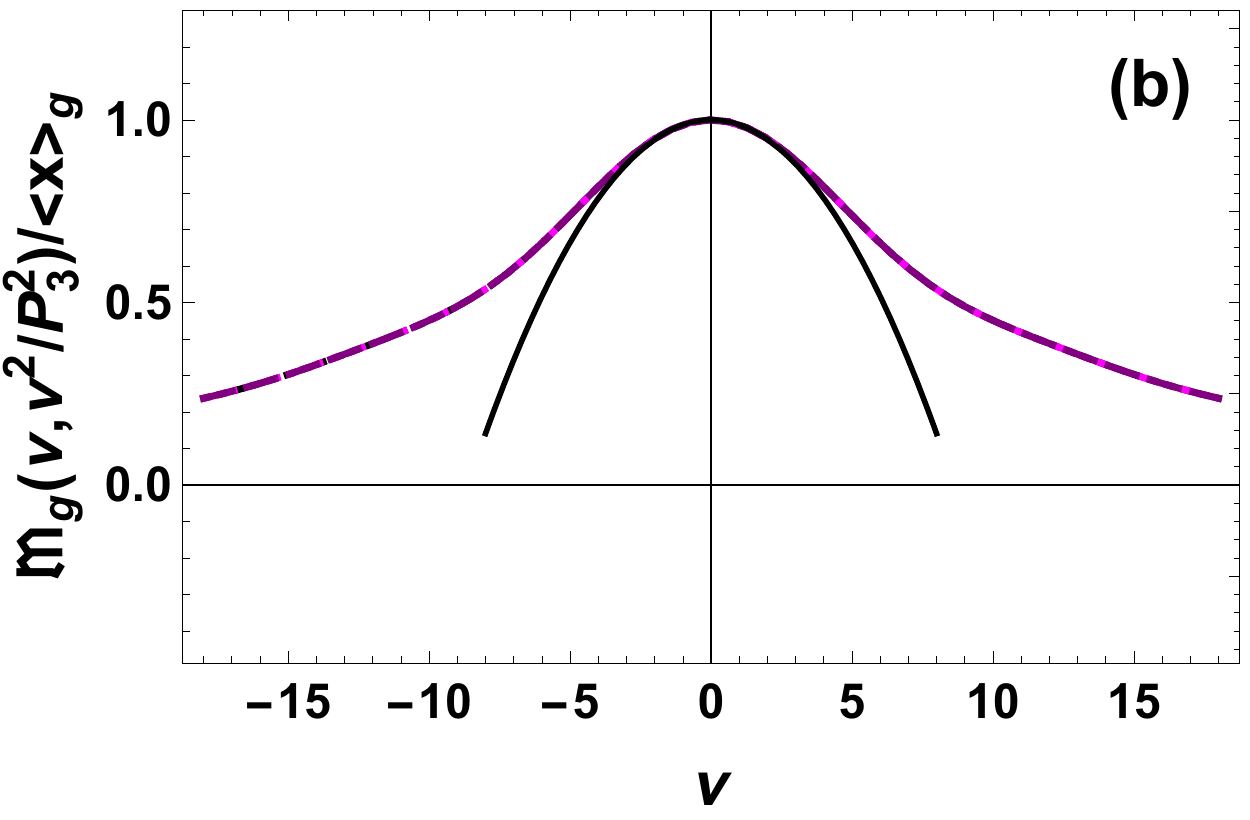}
\end{center}
\vspace{-5mm}
\caption{Same as in Fig.~\ref{fig:Ioffe1} but for the
reduced ITDs of the proton, $\mathfrak{M}$, of Eq.~(\ref{eq:red}). The straight or parabolic solid lines indicate the leading 
expansion at $\nu=0$, as explained in the text. The model curves at various values of $P_3$ overlap, 
displaying universality. \label{fig:ITDred}}
\end{figure}

With the factorization~(\ref{eq:fact}), Eq.~(\ref{eq:b}) becomes the folding formula
\begin{eqnarray} 
\hspace{-7mm} \tilde q(y,P_3)= P_3 \int dx  F[(x-y)P_3] \, q(x) \label{eq:b2}
\end{eqnarray}
of the form factor $F[(x-y)P_3]$ and the PDF. 
Equation~(\ref{eq:b2}) carries a particular ``operational'' simplicity: in the factorized case, QDF is obtained from PDF in terms of a simple folding, which 
washes out the PDF, more and more as $P_3$ is decreased. On the other hand, when $P_3 \to \infty$, the form factor tends to the delta function and 
QDF approaches PDF, in agreement with Eq.~(\ref{eq:limit}).

With the Gaussian form~(\ref{eq:gauss}) one has
\begin{eqnarray} 
\hspace{-7mm} \tilde q(y,P_3)= \frac{1}{\sqrt{2\pi} \Sigma} \int dx  \, e^{-\frac{(x-y)^2}{2 \Sigma^2}}q(x),
\label{eq:b3}
\end{eqnarray}
where 
\begin{eqnarray}
 \Sigma^2 = \frac{1}{\sigma_0^2 P_3^2}=\frac{\langle k_T^2 \rangle}{2P_3^2}. \label{eq:ratio}
\end{eqnarray}
The effective parameter of the mentioned washing-out is thus the ratio $\Sigma^2$ from Eq.~(\ref{eq:ratio}).

In the factorization approximation Eq.~(\ref{eq:TMDzinv2}) becomes 
\begin{eqnarray}
\hspace{-4mm}  h(-P_3 z_3,-z_3^2) =  \hat F(z_3^2) \int {dx}\,e^{i P_3 z_3  x} q(x),  \label{eq:TMDf}
\end{eqnarray}
hence 
\begin{eqnarray}
\hspace{-4mm}  h(0,-z_3^2) =  \hat F(z_3^2)  \label{eq:ff}
\end{eqnarray}
and
\begin{eqnarray}
\hspace{-4mm} \mathfrak{M}(\nu,\nu^2/P_3^2) =  \int {dx}\,e^{i \nu  x} q(x)  \label{eq:ITDf}
\end{eqnarray}
becomes a universal ($P_3$-independent) function.

In the limit of $P_3 \to \infty$, the ITDs also loose the information on the form factor, as
then
\begin{eqnarray}
\hspace{-4mm} h(-\nu,-\nu^2/P_3^2) \to  h(-\nu,0) =  \int {dx}\,e^{i \nu x} q(x), \label{eq:ITDi}
\end{eqnarray}
which gives exactly the same form as Eq.~(\ref{eq:ITDf}).
Note that the form factor $\hat F(z_3^2)$ cancels also from the ratio of the imaginary and real parts, 
\begin{eqnarray}
\frac{{\rm Im} \, h(-P_3 z_3,-z_3^2) }{{\rm Re}\,  h(-P_3 z_3,-z_3^2) } = \frac{\int {dx}\, \sin( P_3 z_3  x) q(x)}{\int {dx}\, \cos( P_3 z_3  x) q(x)}, 
\end{eqnarray}
which also provides a measure of goodness of the factorization ansatz. 

In the factorization ansatz, Eq.~(\ref{eq:deriv2}) takes a simple form, where the width of the transverse-momentum 
distribution of partons is independent of $x$:
\begin{eqnarray}
\hspace{-7mm} 
\left . \frac{d^2}{dz_3^2} h(-P_3 z_3,-z_3^2) \right |_{z_3=0}  = - P_3^2 \langle x^2 \rangle_q - \frac{1}{2}{N_q} \langle k_T^2 \rangle , \label{eq:deriv2f}
\end{eqnarray}
For the gluon distribution analogous results to those listed above are 
immediately obtained.

The remainder of this Section is devoted to an illustration of the derived results in a sample calculation.
We evaluate the matrix elements $h$ and $h_g$ using the 
NNPDF\footnote{We use the file NNPDF30\_nlo\_as\_0118.LHgrid and the interface in Mathematica \cite{Ball:2012cx} for the calculations in this paper.} 
parametrization of the PDFs of the proton 
in the factorization model. As the scale, we take $Q=2.4$~GeV, which corresponds to the lattice spacing of 0.08~fm
used in~\cite{Alexandrou:2015rja,Alexandrou:2016jqi,Alexandrou:2017dzj}.
The factorization ansatz (\ref{eq:fact}) with a Gaussian form factor (\ref{eq:gauss}) is assumed to hold at this scale.
We take $\langle k_T^2 \rangle =0.6~{\rm GeV}^2$ for both the quarks and gluons. 

In Fig.~\ref{fig:Ioffe1} we 
plot the matrix element for the difference of $u$ and $d$ quarks,  $h_{u-d}(-P_3 z_3,-z_3^2)$, 
and $h_g(-P_3 z_3,-z_3^2)$, evaluated at several values of $P_3$ (the values $P_3=0.95$~GeV and 2.4~GeV 
were used in~\cite{Alexandrou:2015rja,Alexandrou:2016jqi,Alexandrou:2017dzj}). The solid line represents the limit of $P_3=0$, where 
$h(0,-z_3^2)=h_g(0,-z_3^2)/\langle x \rangle_g= \hat F(z_3^2)$. We notice clearly the features of Eqs.~({\ref{eq:deriv},\ref{eq:deriv2f}}), with the 
slope of the imaginary parts increasing with $P_3$, and the real parts becoming more and more sharply peaked at the origin. 

Figure~\ref{fig:ioffe2} presents the analogous results for ITDs. Here the solid lines correspond to the $P_3 \to \infty$ limit, i.e., the distributions 
$h(-\nu,0)$ or $h_g(-\nu,0)$ of Eq.~(\ref{eq:ITDi}). We note indeed that as $P_3$ increases, the curves tend to $h(-\nu,0)$ or $h_g(-\nu,0)$, 
but at large values of $\nu$ the convergence is slow.

Finally, in Fig.~\ref{fig:ITDred} we show the
reduced ITDs, which in the factorization ansatz are universal (independent of $P_3$) functions. Note that according to Eqs.~(\ref{eq:ITDf}) and (\ref{eq:ITDi}), 
these are the same curves as the $P_3 \to \infty$ lines from Fig.~\ref{fig:ioffe2}. The straight or parabolic solid lines 
in Fig.~\ref{fig:ITDred} represent the expansion in $\nu$ up to second order, i.e., 
the functions $\nu \langle x \rangle_{u-d}$  and $1+ \frac{1}{2}\nu^2 \langle x \rangle_{u-d}$ for the imaginary and real parts of $h_{u-d}$,
respectively, and the function $1+ \frac{1}{2}\nu^2 \langle x^3 \rangle_{g}/ \langle x \rangle_{g}$ for the case of the gluon distribution.
For the presented NNPDF case, numerically, \mbox{$\langle x \rangle_{u-d}=0.16$}, \mbox{$\langle x^2 \rangle_{u-d}=0.05$}, \mbox{$\langle x \rangle_g=0.44$}, 
and \mbox{$\langle x^3 \rangle_g=0.01$}. Of course, the results conform to the sum rules of Sect.~\ref{sec:sumrules}.

The long tail in the reduced ITDs, prominently seen in Figs.~\ref{fig:Ioffe1} or \ref{fig:ITDred}, is immanently related to the low-$x$ behavior of the 
associated PDFs, which typically involve an integrable singularity as $x \to 0$. For instance, if the distribution behaves low $x$ as $x^{-\alpha}$, with $\alpha<1$, 
(for the moment we use distributions defined in the domain $x \in [0,1]$, which can be converted according to Eq.~(\ref{eq:dom}), then the asymptotic 
behavior of the corresponding ITDs  goes as $\nu^{-1+\alpha}$. 
Note that this long-tail behavior in $\nu$, following from the low-$x$ behavior of the PDFs, 
is inaccessible on the lattice. In contrast, the simulations of~\cite{Orginos:2017kos} 
or~\cite{Alexandrou:2015rja,Alexandrou:2016jqi,Alexandrou:2017dzj} display a rapid fall-off of the matrix elements to zero
around $|\nu|\sim 5$--10. We believe this is associated to the lattice discretization. When $P_3= 2\pi n/L$, with $L$ denoting the longitudinal size and $n$ being
a small natural number, (typically 1--5), then $|\nu|=|P_3 z_3| \le 2\pi n$. This, in turn, via the Fourier transform, sets a lower limit for the accessible 
values of $x$, namely $x>\frac{1}{n}$. 

Having seen that the lattice simulations cannot go to large values of $|\nu|$, a doubt arises concerning the practicality 
of the method. We have demonstrated that the expansion in $\nu$ near the origin, with the coefficients 
given by the $x$-moments of the PDFs, works. Adding some next terms with higher moments would lead to 
further improvement, such that the
expansion would be fairly accurate up to, say, $|\nu|\sim 5$. However, since the ambition of the lattice method based on QDFs is to 
surpass the moment evaluations and provide the  PDFs themselves as functions of $x$ (be it for sufficiently large arguments), one has to verify 
if the``principle of conservation of difficulty'' is possible to circumvent.

\section{QCD evolution and the breaking of factorization \label{sec:break}}

\begin{figure}[tb]
\begin{center}
\includegraphics[angle=0,width=0.45 \textwidth]{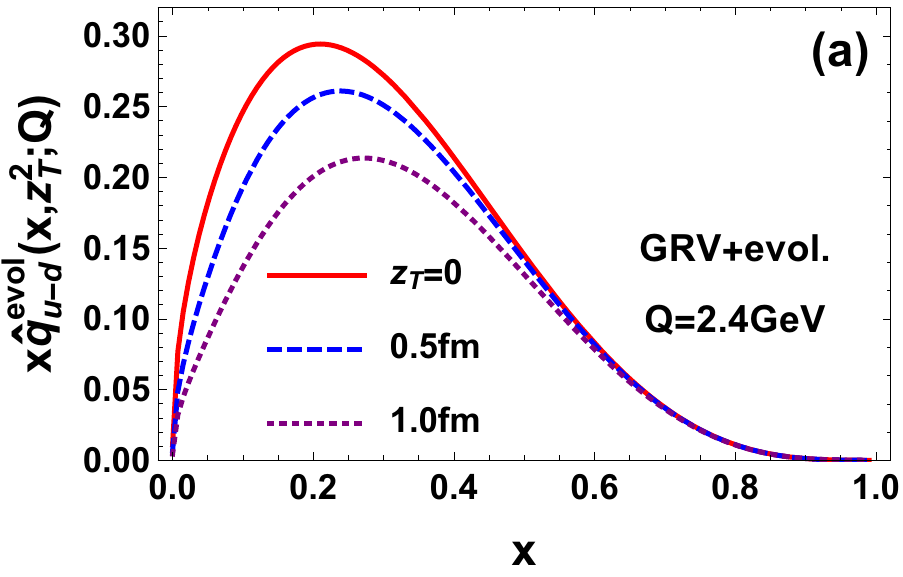}\\
\includegraphics[angle=0,width=0.45 \textwidth]{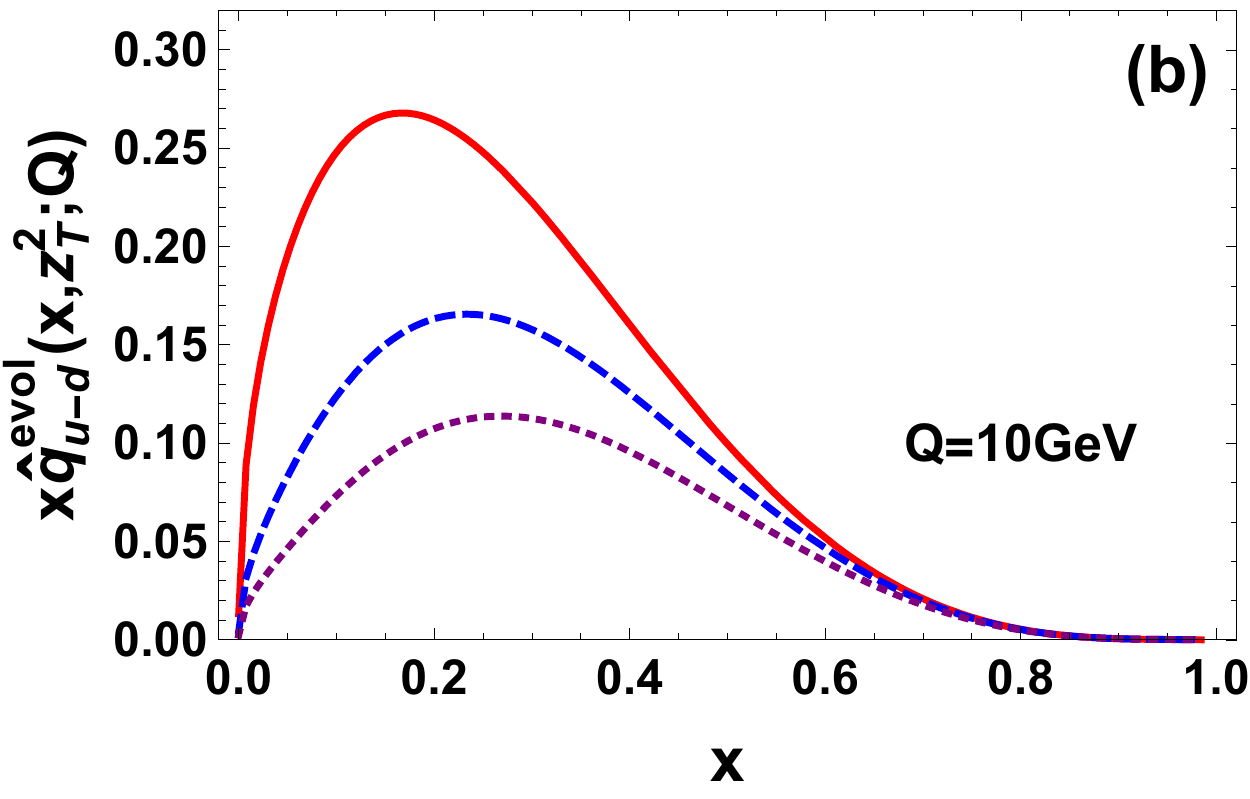}
\end{center}
\vspace{-5mm}
\caption{The $u-d$ TMDs (multiplied with $x$) in the proton, plotted as functions of the momentum fraction 
$x$ at various values of the transverse coordinate $z_T$. The model takes a factorized ansatz at the scale $Q_0=510$~MeV with the GRV parametrization 
and caries out the Kwieci\'nski evolution with Eq.~(\ref{eq:kw}) to the lattice scale (a)~$Q=2.4$~GeV or (b)~\mbox{$Q=10$}~GeV. 
\label{fig:kev}}
\end{figure}

A proper definition of PDFs, QDFs, ITDs, TMDs, etc., requires
specification of the resolution scale, which we generically denote by
$Q$, as it is expected to be the natural choice where the hard scale
is identified with the probing momentum $Q$.  {Here } we treat the
resolution scale as an independent parameter in the problem within the
$\overline{\rm MS}$-renormalization scheme in the continuum, as
opposed to the discrete lattice approach to renormalization. For 
sufficiently fine lattices, the value of the scale can be, roughly speaking, identified
with the lattice spacing expressed in physical units, $a \sim 1/Q$.\footnote{The current limit is $a \sim
0.1~{\rm fm}$, which corresponds to a momentum scale $Q \sim 2~{\rm GeV}$. {This
permits a pQCD matching within the $\overline{\rm
MS}$-renormalization scheme in the continuum.  On the other hand, we recall that the
transverse lattice
approach~\cite{Burkardt:2001mf,Dalley:2002nj,Burkardt:2001jg} with the
resolution scale $1/Q$ corresponding to the transverse lattice spacing,
seems to feature the QCD
evolution in the case of the pion~\cite{Broniowski:2007si}. It also generates a
non-perturbative scale dependence, according to the Wilsonian point of view, which differs 
in that regard from the more popular Euclidean lattice approach.}} 
When $Q$ is large enough, the pQCD approach can be invoked.

A trivial but practically relevant observation is that once we are
able to carry out the QCD evolution for some representation of the
partonic distribution, for instance the TMD, we can then use the
integral transformations unveiled by Radyushkin and spelled out in
Sect.~\ref{sec:def} to effectively carry out the evolution for another
representation, such as QDF.  We can thus rewrite Eq.~(\ref{eq:b})
\begin{eqnarray} 
\hspace{-7mm} \tilde q(y,P_3;Q)= P_3\!\! \int \! dx \!\int \! \frac{dz_2}{2\pi} e^{-i(y-x)z_2 P_3}\hat q(x,z_2^2;Q),
\label{eq:b2n}
\end{eqnarray}
where now the dependence on the scale is explicitly indicated. Our scheme is to evolve the TMD, $\hat q$, and that way 
produce an evolved QDF or ITD. Note that in this treatment $P_3$ is an external (kinematic) variable. 

For the standard unintegrated gluon distribution (or TMD) one has at
hand the Ciafaloni, Catani, Fiorani, and Marchesini (CCFM) evolution
equations~\cite{Ciafaloni:1987ur,Catani:1989yc,Catani:1989sg}, 
which in a sense interpolate between the DGLAP~\cite{Gribov:1972ri,Dokshitzer:1977sg,Altarelli:1977zs} 
and BFKL~\cite{Lipatov:1976zz,Kuraev:1977fs,Balitsky:1978ic} methods.
The CCFM scheme was extended to incorporate quarks by
Kwieci\'nski~\cite{Kwiecinski:2002bx} in the so-called one-loop
approximation. The technicalities standing behind this derivation were
very precisely explained in~\cite{GolecBiernat:2007pu}, see also the
review~\cite{Gustafson:2002jy}, hence we do not give more details
here. 

For our practical purpose it is important we have a ready-to-apply
method with is simple but non-trivial in the present
context.\footnote{One should keep in mind, however, that more
elaborate evolution equations may be needed to account for a specific
gauge-link operator present in the definition of TMDs.}  Moreover,
Kwieci\'nski~\cite{Kwiecinski:2002bx}  showed that in the
transverse-coordinate ($z_T$) representation, the one-loop CCFM
equations become diagonal in $z_T$, possessing the structure very much
like the DGLAP equations for the corresponding integrated parton
distributions (PDFs), but with a modified kernel. For the non-singlet
case they read
\begin{eqnarray}
&& \hspace{-7mm} Q^2{\partial \hat q(x,z_T^2;Q)\over \partial Q^2} =
{\alpha_s(Q^2)\over 2\pi}  \int_0^1d\xi  \,P_{qq}(\xi)
\bigg [ \Theta(\xi-x) \nonumber \\ &&\times  J_0[(1-\xi)Qz_T] \, \hat q\left({x\over \xi},z_T^2;Q\right)
- \hat q(x,z_T^2;Q) \bigg ], \label{eq:kw}
\end{eqnarray}
where $P_{qq}(\xi)$ is the usual splitting function and $J_0$
stands for the Bessel function. The singlet case, embodying the gluon
and sea mixing as well as details and methods of solutions, can be
found
in~\cite{Kwiecinski:2002bx,Gawron:2002kc,Gawron:2003qg,RuizArriola:2004ui}.

The initial condition at the scale $Q_0$ is provided with a factorized form
\begin{eqnarray}
\hat q(x,z_T^2;Q_0)=\hat F(z_T^2) q(x;Q_0), \label{eq:Q0}
\end{eqnarray}
and evolved with Eq.~(\ref{eq:kw}) to the scale $Q$. Since the evolution is diagonal in $z_T$, 
the presence of $\hat F(z_T^2)$ has only a multiplicative effect, 
and the evolved solution has the form
\begin{eqnarray}
\hat q(x,z_T^2;Q)=\hat F(z_T^2) \hat q^{\rm evol}(x,z_T^2;Q). \label{eq:Q}
\end{eqnarray}
In other words, the dependence of the TMD on $z_T$ sits in a factorized trivial component put in by hand, $\hat F(z_T^2)$,\footnote{The 
phenomenological reason to incorporate $\hat F(z_T^2)$ is that without it the obtained width of the $k_T$ distributions seems too narrow.} and 
a dynamically generated non-trivial component, which mixes $z_T$ and $x$, i.e., yields the longitudinal-transverse 
factorization breaking. The
factorization ansatz~(\ref{eq:fact}), which is assumed to hold at a
scale $Q_0$ in Eq.~(\ref{eq:Q0}), is broken at higher scales $Q$.  The breaking increases
with the evolution range and, as we shall see, with decreasing $x$.

In Fig.~\ref{fig:kev} we present the solutions of Eq.~(\ref{eq:kw}) (we plot $\hat q^{\rm evol}$ parts of Eq.~(\ref{eq:Q}), as it 
shows the dynamical effect of the evolution).
For this part of our analysis we take for the PDF the GRV~\cite{Gluck:1998xa} initial
conditions at the scale \mbox{$Q_0=510$}~MeV.\footnote{The reason for using
GRV rather than NNPDF or some other more modern parametrization is that for this
case we have the stored numerical evolution results from
Refs.~\cite{Gawron:2003qg} at hand. Also, the GRV initial scale of 510~MeV is low, which enhances potential 
factorization breaking effects.} For simplicity, we neglect the small effect of the isospin 
asymmetry of the sea quarks. At this scale we use the
factorization formula (\ref{eq:Q0}) with a Gaussian 
form factor and $\langle k_T^2 \rangle_0=0.38~{\rm GeV}^2$. This value is fixed in such a way
that after the evolution to $Q=2.4$~GeV the average width is equal to
the phenomenological number $\overline{\langle k_T^2 \rangle} = 0.57~{\rm GeV}^2$~\cite{Melis:2014pna}.
We note from Fig.~\ref{fig:kev} that an increase of $z_T$ leads to a decrease of the 
distribution, which is accelerated as $Q$ grows. Also, the shape in $x$ is not maintained when 
$z_T$ is changed. This displays the factorization breaking in an explicit manner.

\begin{figure}[tb]
\begin{center}
\includegraphics[angle=0,width=0.45 \textwidth]{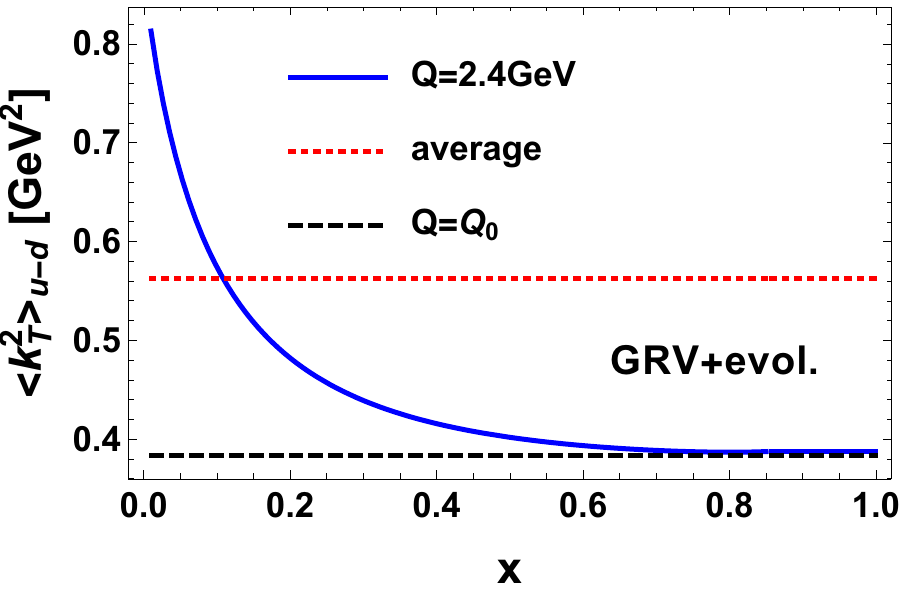}
\end{center}
\vspace{-5mm}
\caption{Transverse-momentum width of the $u-d$ TMD in the proton, plotted as a function of the momentum fraction 
$x$. The model takes a factorized ansatz at the scale $Q_0=510$~MeV
with the GRV parametrization and carries out the Kwieci\'nski
evolution of Eq.~(\ref{eq:kw}) to the lattice scale $Q=2.4$~GeV.  We
notice the broadening of the $k_T$ distribution as $Q$ grows or $x$
decreased. The dashed line indicates $\langle k_T \rangle_0$
originating from the form factor $\hat F$ of Eq.~(\ref{eq:ktdec}),
whereas the dotted line shows the value at $Q$ averaged over $x$.
\label{fig:kt}}
\end{figure}

The evolution of Eq.~(\ref{eq:kw}) leads to a substantial narrowing of the TMDs in $z_T$ or, equivalently, broadening  in $k_T$, 
as $x$ is being decreased. The results for
\begin{eqnarray}
\hspace{-5mm} \langle k_T^2 \rangle_{u-d}(x)=\langle k_T^2 \rangle_0 + \int d^2 k_T \, k_T^2 q^{\rm evol}_{u-d}(x,k_T^2;Q), \label{eq:ktdec}
\end{eqnarray}
after evolution up to $Q=2.4$~GeV, are shown in
Fig.~\ref{fig:kt}. We note a strong dependence on $x$, with $\langle k_T^2 \rangle_{u-d}(x)$ 
growing as $x$ decreases. At $x=1$ there is no effect, which reflects the form of the evolution kernel 
in Eq.~(\ref{eq:kw}). The average width $\overline{\langle k_T^2 \rangle}_{u-d}$
is indicated with a dotted line, whereas the dashed line corresponds to
the value $\langle k_T^2 \rangle_0$ at the scale $Q_0$, following from the assumed 
form factor.  A behavior similar to Fig.~\ref{fig:kt} occurs for other parton species~\cite{Gawron:2003qg}. 

\begin{figure}[tb]
\begin{center}
\includegraphics[angle=0,width=0.45 \textwidth]{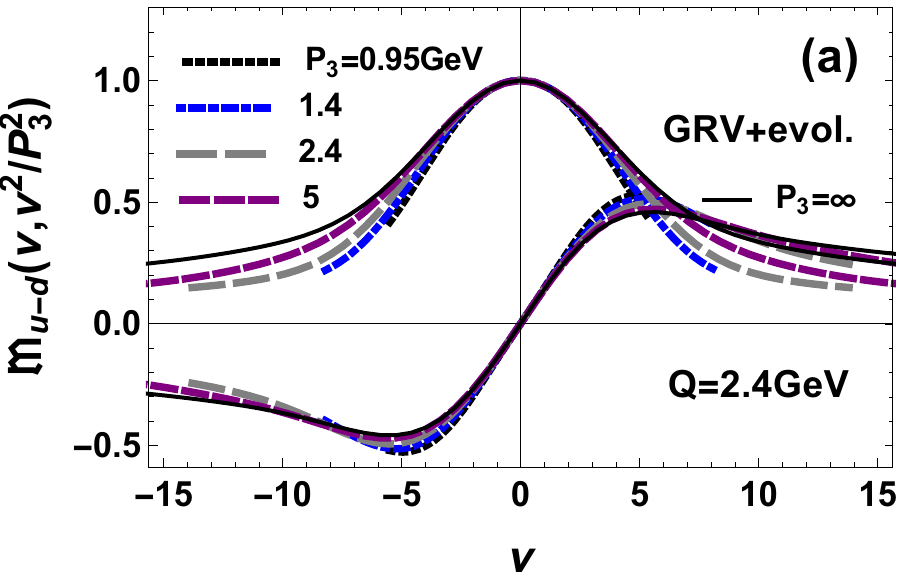} \\ 
\includegraphics[angle=0,width=0.45 \textwidth]{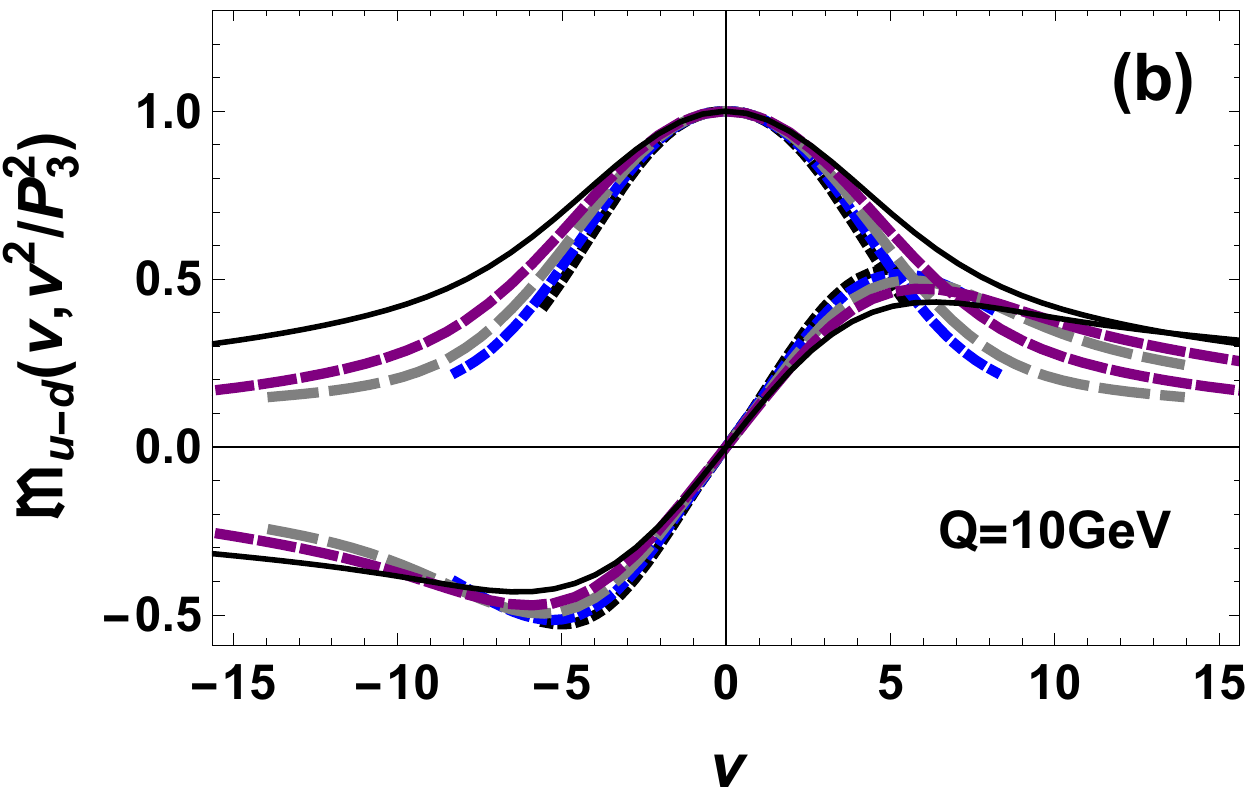} 
\vspace{3mm}
\end{center}
\vspace{-5mm}
\caption{Reduced $u-d$ ITD of the proton at various values of $P_3$ at the evolution scale  (a)~$Q=2.4$~GeV and
(b)~\mbox{$Q=10$}~GeV, obtained from the model described in the text. The solid line represents the $P_3 \to \infty$ limit. \label{fig:ITDred_N_GRV}}
\end{figure}

The key question we wish to address now is whether the described breaking
of the longitudinal-transverse factorization induced by the evolution
of the TMDs leads to noticeable effects in ITDs or QDFs at the scales
relevant for the present-day lattice studies.  We first compare the
results for the reduced $u-d$ ITDs following from the evolved
distributions, which are shown in Fig.~\ref{fig:ITDred_N_GRV}. Recall
that the external form factor effects (i.e., those coming from $\hat
F$) cancel out from this quantity~\cite{Orginos:2017kos}, hence it
serves as a probe for the breaking effects due to evolution. The
dashed curves,\footnote{\label{note1}The curves end at lower values of $|\nu|$
than the range of the plot, which is due to a fixed upper limit for $z_3\simeq
1$~fm in our stored files with evolved TMDs.}  distinguished by
the value of $P_3$, correspond to the model described above, where the
initial condition for the PDF is set at the GRV scale $Q_0=510$~MeV,
and the Kwieci\'nski evolution is carried out to (a) $Q=2.4$~GeV or
(b) $Q=10$~GeV. The solid line shows the $P_3 \to \infty$ case, where
the ITD corresponds to the Fourier transform of the PDF (similarly as the curves in
Fig.~\ref{fig:ITDred}). We note a visible departure from universality, which
at $|\nu|=7$ reaches about 30\% for $Q=2.4$~GeV and 50\% for
$Q=10$~GeV for $P_3 \sim 1$~GeV.

\begin{figure}[tb]
\begin{center}
\includegraphics[angle=0,width=0.45 \textwidth]{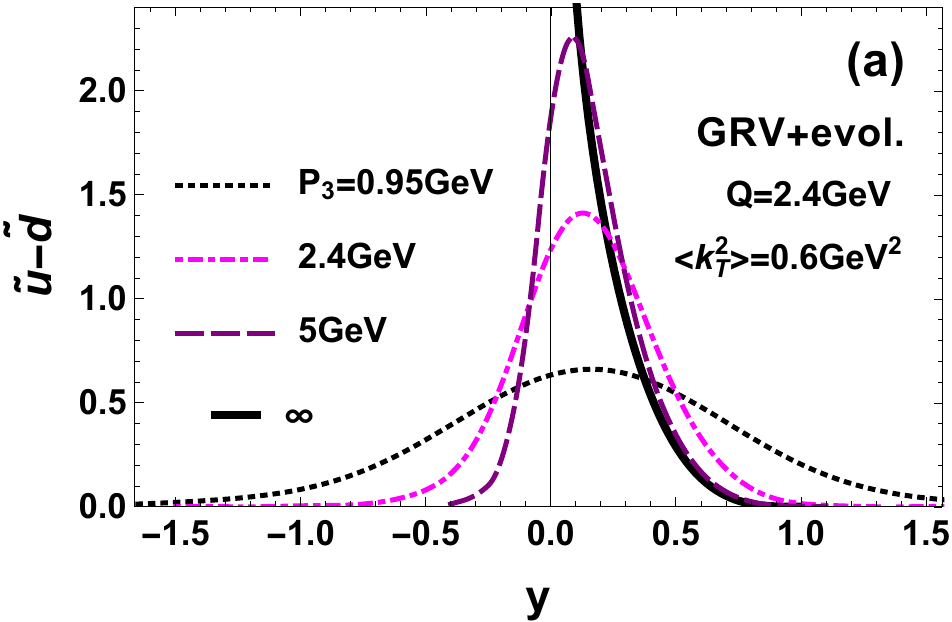} \\ 
\includegraphics[angle=0,width=0.45 \textwidth]{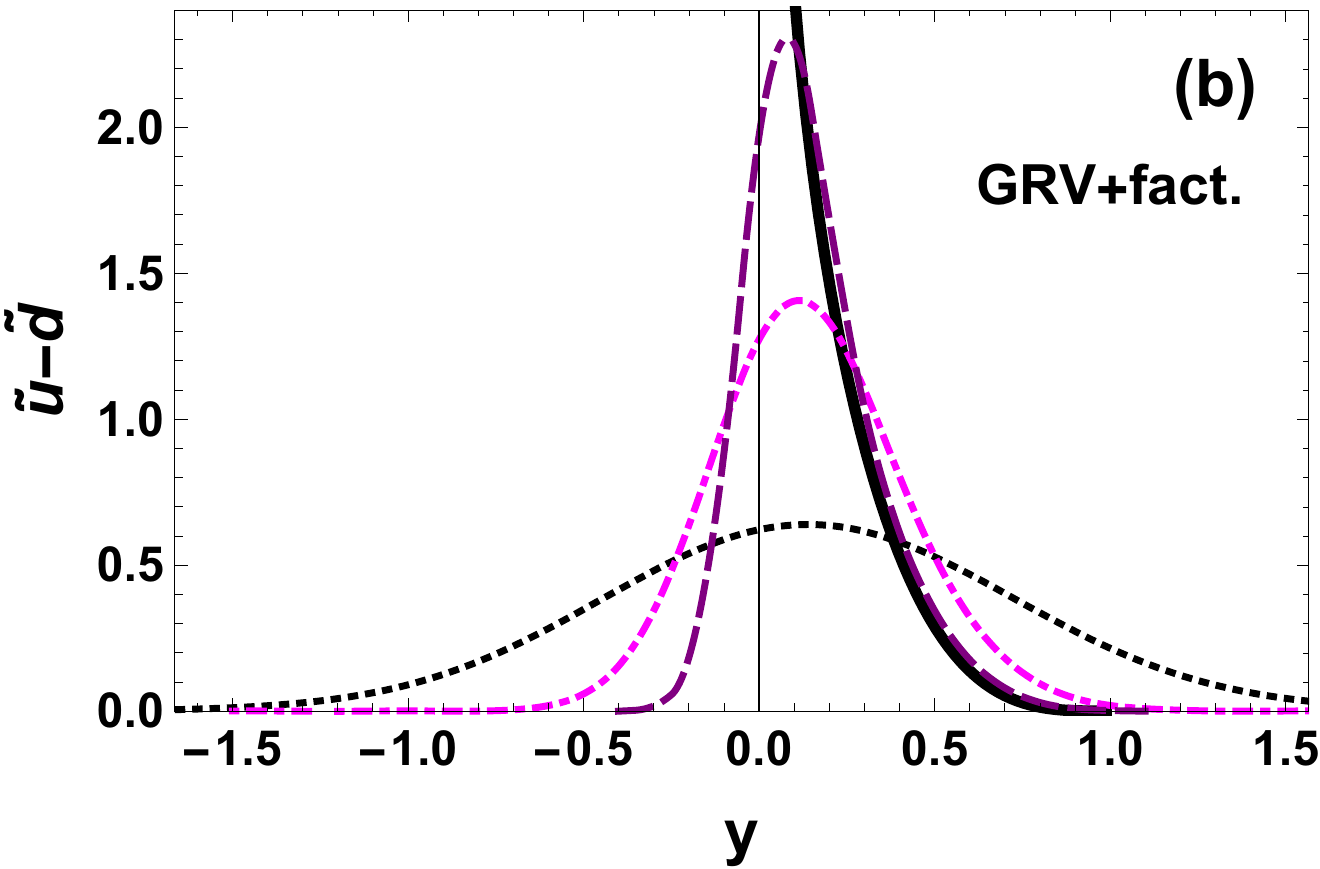} 
\vspace{3mm}
\end{center}
\vspace{-5mm}
\caption{The $\tilde u-\tilde d$ QDF of the proton at several values of $P_3$, obtained from a model with the GRV parameterization of the PDFs 
at $Q_0=510$~MeV, supplied with a Gaussian form factor. In (a) the distributions are evolved to $Q=2.4$~GeV with the Kwieci\'nski 
equations, whereas in  (b) factorization is imposed at the scale $Q$. In both cases the width of the  transverse momentum distribution 
averaged over $x$ is the same and equals $\langle k_T^2 \rangle=0.6~{\rm GeV}^2$.  \label{fig:GRV_QDF}}
\end{figure}

Whereas the factorization breaking effects displayed in
Fig.~\ref{fig:ITDred} seem substantial, or at least relevant at larger
values of $|\nu|$, the issue is to what extent they can influence the
QDFs. The point here is that the form of Eq.~(\ref{eq:b}) leads to
diffusion of the PDF into QDF, which is best seen in the factorized
ansatz (\ref{eq:b2}) or (\ref{eq:b3}). In particular, the PDF at low
values of $x$, where we would expect more effect from factorization
breaking, is diffused more, as the width of the $k_T$ distribution is
larger in that region. As a result, there is no visible effect on the
QDFs from the factorization breaking induces be evolution in our
model. This can be seen from Fig.~\ref{fig:GRV_QDF}, where in
panel~(a) we show the model with the Kwieci\'nski evolution, which
induced the factorization breaking, to be compared with panel (b),
which assumes factorization at the final scale of $Q=2.4$~GeV.  We note that the two cases lead to
essentially identical results.  Thus, as advocated
in~\cite{Orginos:2017kos}, the place to look for potential
factorization breaking are the ITDs and not the QDFs. Our study
supports this conclusion.

\section{Comparison to the Euclidean lattice simulations \label{sec:complat}}

In this Section we compare our results to QDFs obtained from the ETMC full-QCD lattice simulations reported in~\cite{Alexandrou:2016eyt}.
As we have seen that the effects of the transverse-longitudinal factorization seem negligible for QDFs, we return now to the model 
with the NNPDF distributions used in Sect.~\ref{sec:factor} and the simple  Gaussian factorization ansatz (\ref{eq:gauss}) taken 
at the lattice scale $Q=2.4$~GeV. 

\begin{figure}[tb]
\begin{center}
\includegraphics[angle=0,width=0.45 \textwidth]{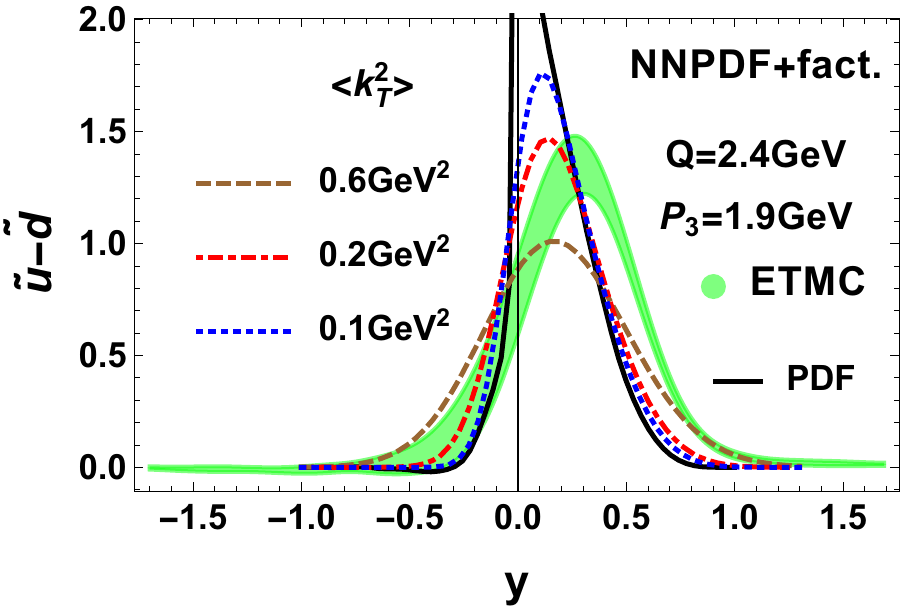}
\end{center}
\vspace{-5mm}
\caption{The $\tilde u - \tilde d$ QDF of the proton in the factorization model with the NNPDF distributions at various values of the width of the $k_T$ 
distribution (lines), compared to the lattice results from ETMC~\cite{Alexandrou:2016eyt} 
(band). Both the model results and the ETMC data are for $P_3=1.9$~GeV. The solid line shows 
the PDF, which is the limit of the QDF at $\langle k_T^2 \rangle \to 0$.
\label{fig:NNPDF_kt}}
\end{figure}

The results for $P_3=1.9$~GeV are shown in Fig.~\ref{fig:NNPDF_kt}, where we use the model with three different values of $\langle k_T^2 \rangle$. We note that 
the model curves move closer to the PDF as $\langle k_T^2 \rangle$ is being decreased, which is obvious from the discussion below Eq.~(\ref{eq:b2}). 
We recall that the combination $\langle k_T^2 \rangle/P_3^2$ is the relevant parameter, and its going to zero provides the PDF limit.
At the same time, the comparison to the ETMC data, represented with a band, is qualitative only, except perhaps the large-$y$ region for 
$\langle k_T^2 \rangle=0.6~{\rm GeV}^2$.

Figure~\ref{fig:NNPDF_kt} presents a similar study, where we keep  $\langle k_T^2 \rangle$ at the value of $0.3~{\rm GeV}^2$~\cite{Bacchetta:2017gcc}, but 
change the value of $P_3$. Comparison is made to the corresponding 
three QDF extractions from the ETMC data, indicated with the bands. Again, the model curves are 
substantially away from the lattice extractions.

\begin{figure}[tb]
\begin{center}
\includegraphics[angle=0,width=0.45 \textwidth]{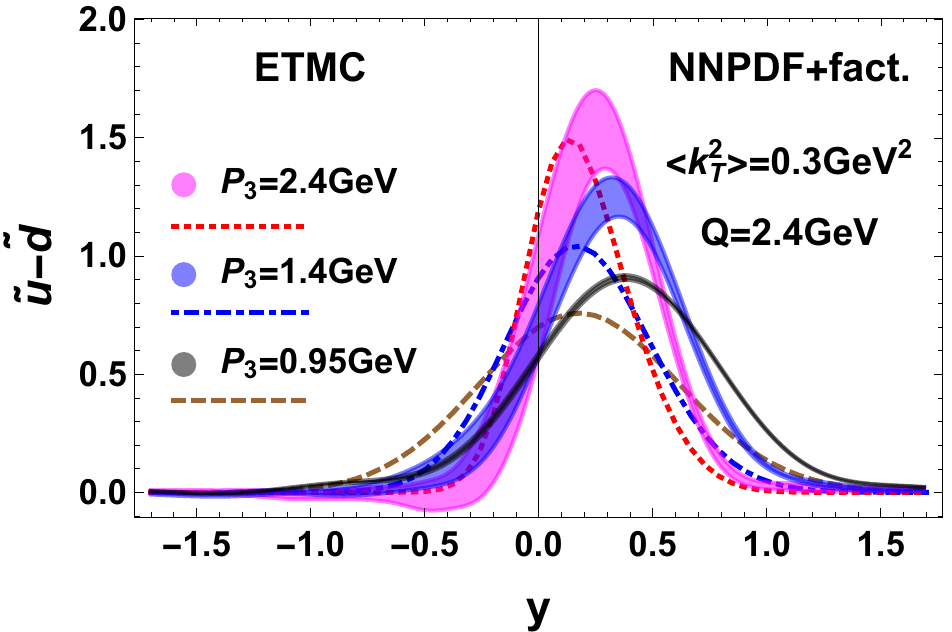}
\end{center}
\vspace{-5mm}
\caption{Same as in Fig.~\ref{fig:NNPDF_kt}, but for the case where $\langle k_T^2 \rangle=0.3~{\rm GeV}^2$ is fixed and
$P_3$ changed.
\label{fig:NNPDF_P3}}
\end{figure}

There are several possible reasons for the discrepancy. First, as
discussed in Appendix~\ref{app:decomp}, the extraction of QDF
in~\cite{Alexandrou:2015rja,Alexandrou:2016eyt,Alexandrou:2016jqi,Alexandrou:2017dzj}
uses a prescription retaining the structure proportional to
$z^\mu$. Then, the Radyushkin QDF-TMD relation (\ref{eq:rad}) receives
corrections subleading in the twist expansion.  Moreover, this choice
leads to mixing with a subleading-twist scalar channel which needs to
be disentangled \cite{Alexandrou:2017huk}. Another issue is the value
of the pion mass, which in the ETMC simulations is
$m_\pi=370$~MeV. One artifact, possibly caused by a large departure
from the physical pion mass limit, is a large value of the momentum
fraction $\langle x \rangle_{u-d}=0.23$ (cf. Table~I
of~\cite{Alexandrou:2016jqi}), compared to the phenomenological value
of 0.16. Thus, quite naturally, the lattice QDFs are moved to the right from
the PDF, as in Figs.~\ref{fig:NNPDF_kt} and~\ref{fig:NNPDF_P3}.  A
proper extrapolation in $m_\pi$ down to physical value may resolve
this problem.  
The target-mass corrections~\cite{Alexandrou:2015rja,Radyushkin:2017ffo} also move the lattice extractions closer to the 
data. Apart from the issues mentioned above, there are also
typical lattice problems, such as a finite cut-off from the lattice
spacing, volume effects, the source-sink separation, etc.

We note that the quenched simulation in~\cite{Orginos:2017kos}, which served as a proof of concept of the 
invented methods and where the $P^0$ projection discussed in Appendix~\ref{app:decomp} was used, 
the value of the pion mass was 600~MeV.
In this study, the PDF extracted from the lattice is also visibly to the right of the phenomenological distribution.  

Besides these issues, we note from Fig.~\ref{fig:NNPDF_P3} that the needed values for $P_3$ to achieve a few-percent agreement with the PDF limit for 
$x>0.15$ are $P_3>5~{\rm GeV}$, or more appropriately, $\langle k_T^2 \rangle/P_3^2 < 0.025$.

\begin{figure}[tb]
\begin{center}
\includegraphics[angle=0,width=0.45 \textwidth]{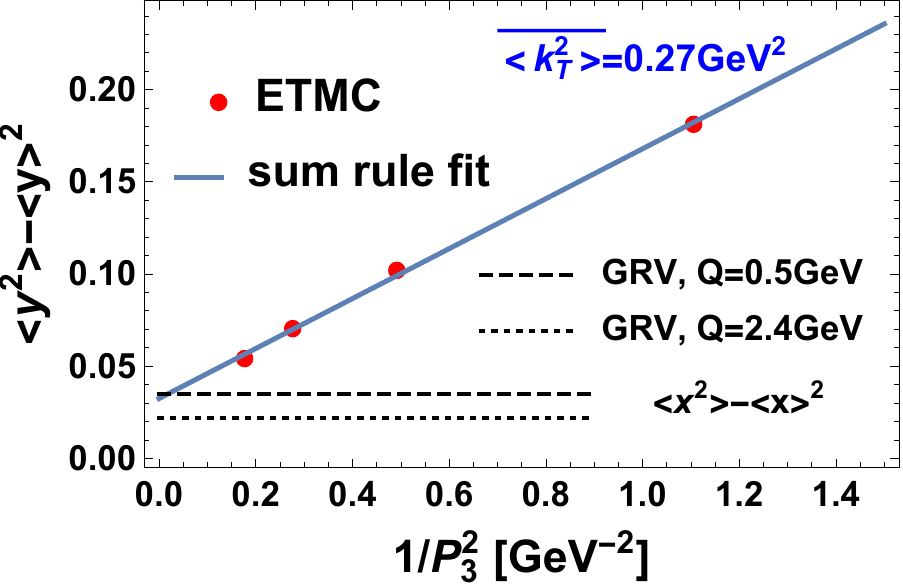}
\end{center}
\vspace{-5mm}
\caption{The sum rule of Eq.~\ref{eq:cen} at work. We use the lattice data 
from the ETMC collaboration~\cite{Alexandrou:2016eyt} to compute the
second $y$-moment of the QDFs. We note that the value at the origin gives
the second central $x$-moment of the PDF. The horizontal lines correspond 
to the phenomenological GRV analysis~\cite{Gluck:1998xa} for the values
$Q=0.5$ and $2.4~{\rm GeV}$. The slope yields the value of $\overline{\langle k_T^2 \rangle}_{u-d} = 0.27~{\rm GeV}^2$
for the spread of the transverse momentum distribution. \label{fig:sr}}
\end{figure}

{Finally, we illustrate in the nucleon case the sum rules
discussed in Section~\ref{sec:sumrules}, which for the second central\footnote{We use central moments here
to avoid problems die to the fact that the mean $x_{u-d}$ is too large compared to phenomenological  
parameterizations.} moment  (\ref{eq:deriv},\ref{eq:deriv2}) yield}
\begin{eqnarray} 
\langle y^2 \rangle - \langle y \rangle^2 = \langle x^2 \rangle - \langle x \rangle^2 + \frac{\overline{\langle k_T^2 \rangle}}{2P_3^2}. \label{eq:cen}
\end{eqnarray}
{This relation allows us to extract the TMD width, $\overline{\langle
k_T^2 \rangle}$, directly from the lattice data on QDFs from the ETMC
collaboration~\cite{Alexandrou:2016eyt}.\footnote{The point at $P_3=0.95$~GeV is obtained for the 
Gaussian smearing data, and the remaining points from the momentum smearing data.}
We just make a linear fit  
of the form $A + B/P_3^2$. The result
is depicted in Fig.~\ref{fig:sr}, where a clear straight line can be
seen. The slope yields the value of $\langle k_T^2 \rangle_{u-d} = 0.27~{\rm GeV}^2$.~\footnote{The
numerical resemblance with SQM model calculations of the pion, yielding $\langle
k_T^2 \rangle = m_\rho^2/2$, is worth
noticing~\cite{RuizArriola:2003wi,RuizArriola:2003bs}}  Another 
determination of this quantity was made in the lattice study~\cite{Musch:2010ka} 
by means of a Gaussian fit in $k_T$, with the result 
$\langle k_T^2 \rangle_{u-d} = (0.16(1)~{\rm GeV})^2$ at $m_\pi=600$~MeV.
In addition, we note from Fig.~\ref{fig:sr} an agreement of the second central $x$ moment with the
phenomenological GRV analysis~\cite{Gluck:1998xa}, holding in the range
$Q=0.5-2.4 {\rm GeV}$, with a better agreement for the lower scale.}

\section{Predictions for the pion \label{sec:pion}}

Finally, we wish to make some predictions for the pion, which undoubtedly also will be soon analyzed on the lattice
in the context of ITDs or QDFs. Note that a similar object, namely the pion quasi-distribution amplitude~\cite{Ji:2013dva,Radyushkin:2017gjd}, has been evaluated 
on the lattice~\cite{Zhang:2017bzy} and reproduced favorably in a chiral quark model~\cite{Broniowski:2017wbr}.

The phenomenological parton distributions for the pion were extracted from the Drell-Yan and the prompt
photon emission experiments. The parametrization provided in~\cite{Sutton:1991ay},
denoted as SMRS (see Table VII, NA10 case at $Q^2 = 5~{\rm GeV}^2$), reads 
\begin{eqnarray}
V_\pi(x) = A x^{-0.4} (1 - x)^{1.08}, \label{eq:SMRS}
\end{eqnarray}
for the valence quark PDF of the pion.  We use his form to derive, with the
techniques of the previous Sections, the corresponding QDF and the
reduced ITD.

\begin{figure}[tb]
\begin{center}
\includegraphics[angle=0,width=0.45 \textwidth]{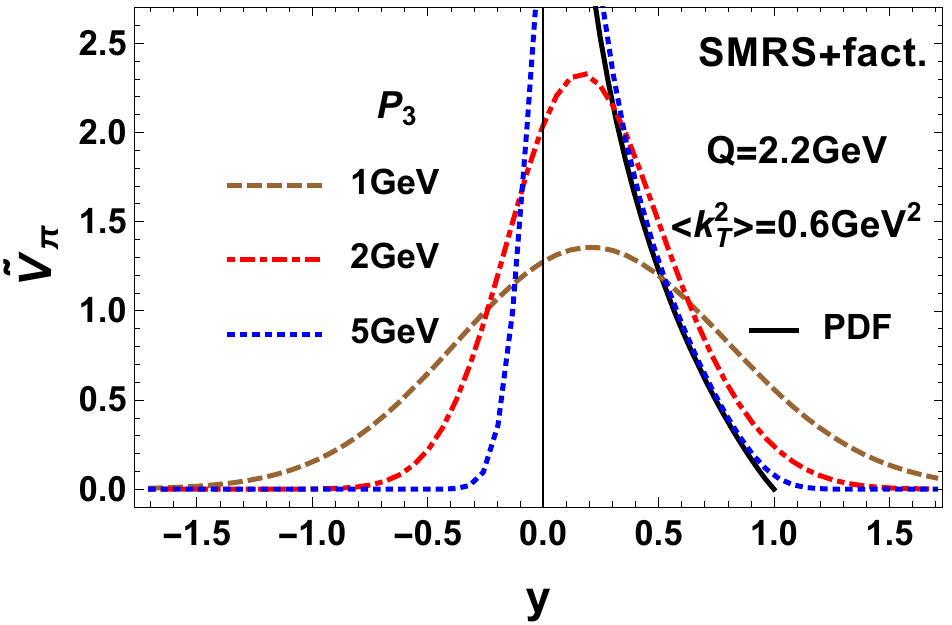}
\end{center}
\vspace{-5mm}
\caption{Valence QDFs of the pion at various values of $P_3$ for the SMRS parametrization~(\ref{eq:SMRS}). Factorization ansatz is 
imposed at the scale $Q=2.2$~GeV. The solid line indicates the valence PDF of the pion.
\label{fig:SMRS}}
\end{figure}

In Fig.~\ref{fig:SMRS} we show the valence QDFs of the pion, $V_\pi$,  at several values of $P_3$ in a model, where a Gaussian factorization ansatz of width 
$\langle k_T^2 \rangle=0.6~{\rm GeV}^2$ is imposed at the 
SMRS scale $Q=2.2$~GeV, with the PDF taken from Eq.~(\ref{eq:SMRS}). We note a behavior qualitatively similar to the proton case of 
Fig.~\ref{fig:GRV_QDF}, with the QDF converging to within a few percent to the PDF at $P_3>5$GeV (for $x>0.15$). 

We have also carried out a similar analysis with  the factorization breaking in the pion due to the Kwieci\'nski evolution starting
from the GRS~\cite{Gluck:1999xe} parametrization at the scale of $Q_0=510$~MeV and carried out up to $Q=2.2$~GeV, and found small 
factorization breaking effects in QDFs, similarly to the proton case discussed in detail in Section~\ref{sec:break}.

The longitudinal-transverse factorization breaking due to the QCD evolution naturally increases with 
the evolution ratio $r\equiv \alpha_{\rm QCD}(Q_0)/\alpha_{\rm QCD}(Q)$. 
Thus the effect will be enhanced in approaches where $r$ is large. This is notoriously the case of the chiral quark models ($\chi$QM)
(for a review in the context of PDF and PDA analyses, see~\cite{RuizArriola:2002wr} and references therein), where 
the quark-model scale $Q_0$ is very low, $Q_0\sim 320$~MeV, and $r\simeq 7$ for $Q=2.2$~GeV.  The quark-model scale is defined 
as the scale where the valence quarks, which are the only degrees of freedom in the model, saturate the momentum sum rule. 

\begin{figure}[tb]
\begin{center}
\includegraphics[angle=0,width=0.45 \textwidth]{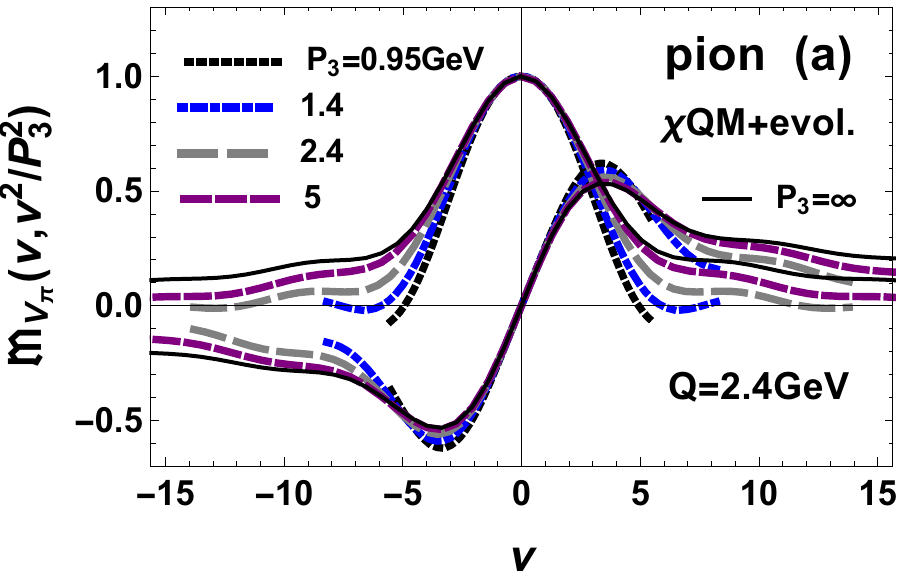} \\ 
\includegraphics[angle=0,width=0.45 \textwidth]{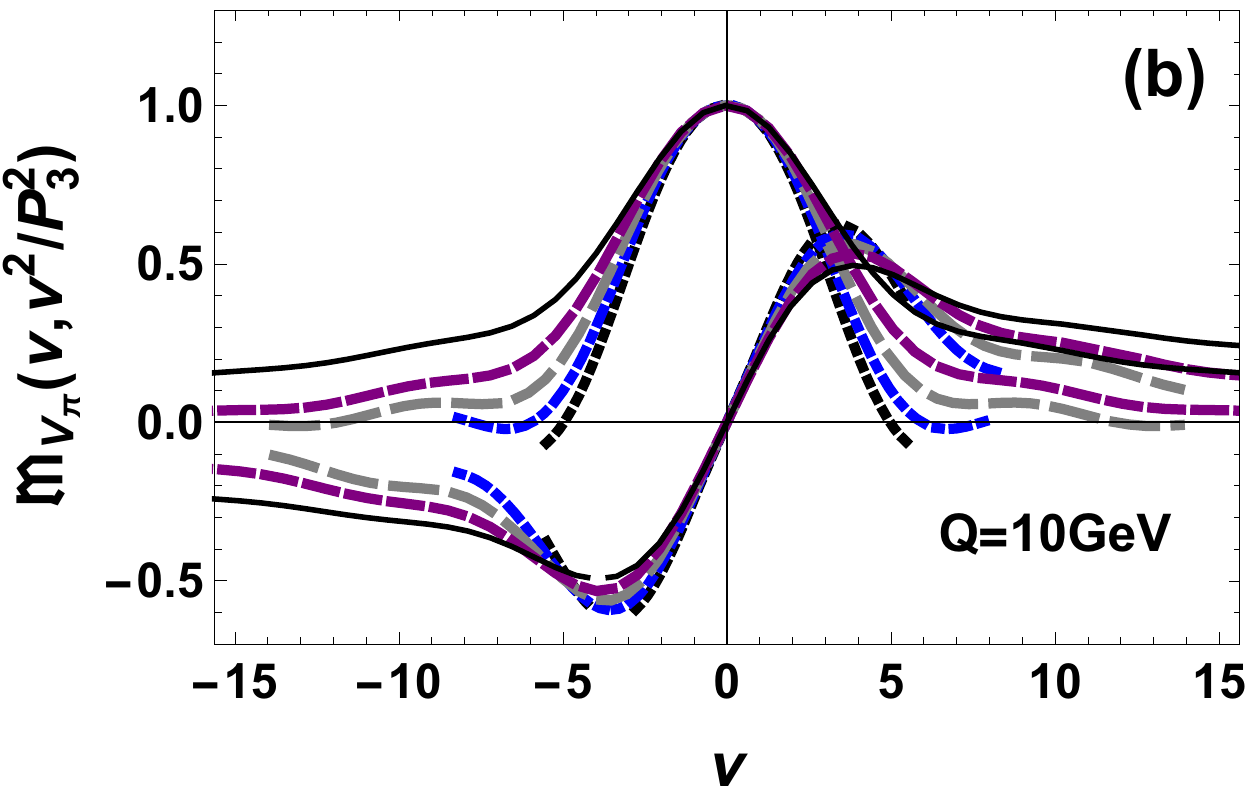} 
\vspace{3mm}
\end{center}
\vspace{-5mm}
\caption{
Reduced valence ITD of the pion, evaluated in the Chiral Quark Model ($\chi$QM) at various values of $P_3$. 
At the origin, the real and imaginary parts equal 1 or 0, respectively. 
\label{fig:ITDred_pion}}
\end{figure}

In Fig~\ref{fig:ITDred_pion} we present the reduced valence ITD of the pion, evaluated in $\chi$QM, where the PDF at the 
initial scale $Q_0$ has a constant value~\cite{Davidson:1994uv}, and the Kwieci\'nski evolution~(\ref{eq:kw}) is performed up to $Q=2.2$~GeV.
We notice strong violation effects, larger than for the analogous plot for the nucleon~(\ref{fig:ITDred_N_GRV}), which is a result of an 
increased evolution ratio $r$. We note that at $|\nu|=5$ the effect reaches 100\% for the 
lower values of $P_3$.\footref{note1}

\section{Conclusions}

The {\em ab initio} determination of the parton distribution functions
is a formidably complex problem which remains a pending issue in
hadronic structure. Whereas the $x$-moments method has been for a long
time the only available scheme for Euclidean lattices, the QDF methodology
proposed by Ji has opened a new venue in the field by considering
space-like correlators boosted to a finite momentum, and eventually
extrapolating to the infinite momentum limit. These apparently
auxiliary new mathematical objects have been found by Radyushkin to be
intertwined with the well known TMDs, or more generally, with the 
pseudo-distributions. This makes QDFs at {\em finite} longitudinal momentum
interesting on their own. As a bonus, this connection suggests a
working scheme to implement the QCD evolution for QDFs via an evolution
of TMDs, which has been studied for many years, offering working
prescriptions ready to use.

In the present paper we have profited from the Radyushkin relation
between the QDFs and TMDs or ITDs in several ways. First, we have
written down some useful sum rules which can be easily used as
consistency checks for the lattice studies. The sum rules show that at
low values of the Ioffe time, the reduced ITDs are essentially
dominated with the lowest $x$-moments of PDFs. Application of the sum
rules to ITDs also allows one, with sufficiently accurate lattice
data, for an extraction of the transverse-momentum widths of TMDs.
{We have checked favorably the lowest sum rule on the ETMC
lattice data and obtained the $k_T$-width of the TMD of
the nucleon at a low scale.
}

Second, we have conducted a phenomenological analysis of the QCD
evolution effects on the quark and gluon components of the proton
using the Kwieci\'nski extension of the one-loop CCFM equations. Our
method uses the established parameterizations of PDFs in conjunction
with the widely employed longitudinal-transverse factorization
ansatz imposed at a low momentum scale. 
We have focused on the examination of the factorization
breaking due to the QCD evolution.
While, strictly
speaking, the factorization ansatz can only hold at a given reference scale, we
have shown that the breaking of factorization is not numerically very large as
long as the evolution ratio is not large. Whereas the breaking is visible in the reduced ITDs, 
it essentially disappears from QDFs at the presently available scales.   
This finding is in agreement with
factorization studies on the lattice,
where factorization is found to hold in a relatively wide range. The
reason is due to a rather weak effect of the QCD evolution at the scales presently available on the lattice.
All these results make the {\it a priori} naive
but actually valid factorization property even more intriguing from a theoretical point
of view.

Finally, we have presented predictions for the valence-quark QDF in the pion, as well as for the 
corresponding reduced ITD. To enhance the possible effects of the longitudinal-transverse 
factorization breaking, we have used chiral quark models, where the QCD evolution ration is large 
and the effect are largely enhanced. This calculation may serve as a limit of how large the breaking effects
could be.

\begin{acknowledgments}
We are very grateful to Krzysztof Cichy for providing the data points
from the European Twisted Mass Collaboration used in the figures, and
for numerous valuable discussions. We also thank Anatoly Radyushkin 
for comments on the paper. 
This work was supported by the Polish National
Science Center grant 2015/19/B/ST2/00937, by the Spanish Mineco (Grants FIS2014-59386-P and FIS2017-85053-C2-1-P), and by the
Junta de Andaluc\'{\i}a (grant FQM225-05), 
\end{acknowledgments}

\begin{appendix}

\section{Decomposition of the matrix element \label{app:decomp}}

Rewriting Eq.~(\ref{eq:fdef}) for brevity as
\begin{eqnarray}
M^\mu = P^\mu A + z^\mu B,
\end{eqnarray}
we find from contractions with $P_\mu$ and $z_\mu$ the relations
\begin{eqnarray}
&& A=\frac{M\cdot z \, P\cdot z- M \cdot p \, z^2}{P \cdot z^2-P^2 z^2}, \nonumber \\ 
&& B=\frac{M\cdot p \, P \cdot z- M \cdot z \, P^2}{P \cdot z^2-P^2 z^2}. \label{eq:dec}
\end{eqnarray}
We may now consider the kinematic cases of interest.
For PDFs,  the only nonzero component of $z$ is $z^-$, hence taking $\gamma^+$ in the definition~(\ref{eq:fdef}) yields 
\begin{eqnarray}
M^+ = P^+ A. 
\end{eqnarray}
The same relation holds for TMDs, where $z^-$ and $z_T$ are nonzero. 
For the kinematics of QDFs defined by Ji~\cite{Ji:2013dva}, only $z^3$ is nonzero, and 
\begin{eqnarray}
M^3 = P^3 A+z^3 B, \label{eq:mix}
\end{eqnarray}
where both $A$ and $B$ structures enter, precluding a generic link to TMD, which contains $A$ only.
In Ref.~\cite{Orginos:2017kos} it is proposed to take 
\begin{eqnarray}
M^0 = P^0 A.
\end{eqnarray}
Note that despite the mixing in Eq.~(\ref{eq:mix}), in the limit of $P_3 \to \infty$ (under assumptions of regularity of $B$), the term with $A$ dominates, 
hence the asymptotic link to the PDF follows.

We note that in~\cite{Alexandrou:2015rja,Alexandrou:2016eyt,Alexandrou:2016jqi,Alexandrou:2017dzj} the $M^3$ prescription is used, 
hence the above difficulty arises. Moreover, this choice leads to mixing of the unpolarized QDF with the twist-3 scalar correlator~\cite{Alexandrou:2017huk}, 
adding to technical difficulties. 

One could also use the prescription with $M^3$, but with $z$ having
only a non-vanishing transverse component, $z_2$. In that case $M^3 =
P^3 A$.

\section{Transversity relation for the pion wave function \label{sec:tr}} 

Consider  the relation~\cite{Broniowski:2009dt}
\begin{eqnarray}
\Psi_a(P \cdot z , z^2) = \int_0^1 d\alpha e^{i (2\alpha-1) P \cdot z}\Phi_a ( \alpha, z^2 ), \label{eq:Phi}
\end{eqnarray}
where $\Psi_a(z\cdot q, z^2)$ is the pion wave function (related to the Bethe-Salpeter amplitude in the given tensor channel $a$), 
and $\Phi_a(z\cdot q, z^2)$ is its Fourier transform. The functions, as Lorentz invariants, depend on the two available scalars $P \cdot z$ and $z^2$.
Choosing two specific frames: equal-time (ET), 
with $z=(0,\vec{r})$ and $P=(m_\pi, 0)$, and the infinite-momentum light-cone frame (LC), with $z_+=0$ and 
$P=(P_0,0,0,P_3)=\lim_{P_3 \to \infty}(\sqrt{m_\pi^2+P_3^2}, 0,0,P_3)$, hence $P_+ z_-=P \cdot z=0$,
one derives a relation between the ET and LC pion wave functions
\begin{eqnarray}
 \Psi^{\rm ET}_a(0, -r^2) = \int_0^1 d\alpha \Phi^{\rm LC}_a (\alpha, -r^2 ). \label{eq:ETLC}
\end{eqnarray}
The integration variable $\alpha$ in Eq.~(\ref{eq:ETLC}) acquires the meaning of the light-cone momentum fraction of the pion carried by the quark.

We bring up this example, since the discussion in this paper concerning the distribution functions bears a lot of similarity. 
In that case, direct analogs of $\Phi_a( \alpha, z^2 )$ are the {\em pseudo-distributions} introduced by
Radyushkin~\cite{Radyushkin:2017cyf}.

\section{Derivation of the Radyushkin relation \label{sec:rad}} 

In this Appendix 
we present, for completeness, a pedestrian derivation of Eq.~(\ref{eq:rad}), which is
based solely on the Lorentz invariance~\cite{Radyushkin:2017cyf} of
the matrix element $h$ appearing in the decomposition~(\ref{eq:fdef}).

In the definition of TMD we encounter, by construction, the matrix element
\begin{eqnarray} 
h(P\cdot z, z^2)|_{z_+=0}= h(P_+ z_- , -z_1^2-z_2^2), \label{eq:lor1}
\end{eqnarray}
whereas in QDF
\begin{eqnarray}
h( P\cdot z, z^2)|_{z_0=0, z_1=z_2=0}= h(-P_3 z_3 , -z_3^2). \label{eq:lor2}
\end{eqnarray}
Now, following~\cite{Radyushkin:2017cyf}, one takes the specific value 
\begin{eqnarray}
k_2=(x-y)P_3
\end{eqnarray} 
in the definition~(\ref{eq:TMD}). Then, using Eq.~(\ref{eq:TMD}) and 
carrying out the two integrations from Eq.~(\ref{eq:rad}) we readily find
\begin{eqnarray}
&&\int \!\!dk_1 \!\!\int \!\!dx \,  q(x,k_1,(y-x)P_3) = P^+ \!\!\! \int dz^- \delta(P_+ z_- +P_3 z_2) \nonumber \\
&& ~~~ \times \int dz_1 \delta(z_1) \!\!\!\int \frac{dz_2}{2\pi} e^{-i y P_3 z_2}  h(P_+ z_-,-z_1^2-z_2^2) = \nonumber \\
&&  ~~~\int \frac{dz_2}{2\pi} e^{-i y P_3 z_2}  h(-P_3 z_2,-z_2^2) =\nonumber \\ 
&&  ~~~\int \frac{dz_3}{2\pi} e^{-i y P_3 z_3}  h(-P_3 z_3,-z_3^2)  \equiv \frac{1}{P_3} \tilde q(y,P_3). \label{eq:quasi}
\end{eqnarray}
Since the support of $q(x,k_T)$ is $x\in [-1,1]$, the $x$ integration
can be formally carried in $(-\infty,\infty)$, yielding the delta
function.  In the last line we have changed the {\em notation} for the
dummy integration variable, $z_2 \to z_3$, which finally yields
Eq.~(\ref{eq:rad}).

\end{appendix}

\bibliography{QDF-TMD}

\begin{thebibliography}{78}%
\makeatletter
\providecommand \@ifxundefined [1]{%
 \@ifx{#1\undefined}
}%
\providecommand \@ifnum [1]{%
 \ifnum #1\expandafter \@firstoftwo
 \else \expandafter \@secondoftwo
 \fi
}%
\providecommand \@ifx [1]{%
 \ifx #1\expandafter \@firstoftwo
 \else \expandafter \@secondoftwo
 \fi
}%
\providecommand \natexlab [1]{#1}%
\providecommand \enquote  [1]{``#1''}%
\providecommand \bibnamefont  [1]{#1}%
\providecommand \bibfnamefont [1]{#1}%
\providecommand \citenamefont [1]{#1}%
\providecommand \href@noop [0]{\@secondoftwo}%
\providecommand \href [0]{\begingroup \@sanitize@url \@href}%
\providecommand \@href[1]{\@@startlink{#1}\@@href}%
\providecommand \@@href[1]{\endgroup#1\@@endlink}%
\providecommand \@sanitize@url [0]{\catcode `\\12\catcode `\$12\catcode
  `\&12\catcode `\#12\catcode `\^12\catcode `\_12\catcode `\%12\relax}%
\providecommand \@@startlink[1]{}%
\providecommand \@@endlink[0]{}%
\providecommand \url  [0]{\begingroup\@sanitize@url \@url }%
\providecommand \@url [1]{\endgroup\@href {#1}{\urlprefix }}%
\providecommand \urlprefix  [0]{URL }%
\providecommand \Eprint [0]{\href }%
\providecommand \doibase [0]{http://dx.doi.org/}%
\providecommand \selectlanguage [0]{\@gobble}%
\providecommand \bibinfo  [0]{\@secondoftwo}%
\providecommand \bibfield  [0]{\@secondoftwo}%
\providecommand \translation [1]{[#1]}%
\providecommand \BibitemOpen [0]{}%
\providecommand \bibitemStop [0]{}%
\providecommand \bibitemNoStop [0]{.\EOS\space}%
\providecommand \EOS [0]{\spacefactor3000\relax}%
\providecommand \BibitemShut  [1]{\csname bibitem#1\endcsname}%
\let\auto@bib@innerbib\@empty
\bibitem [{\citenamefont {Collins}(2013)}]{Collins:2011zzd}%
  \BibitemOpen
  \bibfield  {author} {\bibinfo {author} {\bibfnamefont {J.}~\bibnamefont
  {Collins}},\ }\href@noop {} {\emph {\bibinfo {title} {{Foundations of
  perturbative QCD}}}}\ (\bibinfo  {publisher} {Cambridge University Press},\
  \bibinfo {year} {2013})\BibitemShut {NoStop}%
\bibitem [{\citenamefont {Burkardt}\ and\ \citenamefont
  {Seal}(2002)}]{Burkardt:2001mf}%
  \BibitemOpen
  \bibfield  {author} {\bibinfo {author} {\bibfnamefont {M.}~\bibnamefont
  {Burkardt}}\ and\ \bibinfo {author} {\bibfnamefont {S.~K.}\ \bibnamefont
  {Seal}},\ }\href {\doibase 10.1103/PhysRevD.65.034501} {\bibfield  {journal}
  {\bibinfo  {journal} {Phys. Rev.}\ }\textbf {\bibinfo {volume} {D65}},\
  \bibinfo {pages} {034501} (\bibinfo {year} {2002})},\ \Eprint
  {http://arxiv.org/abs/hep-ph/0102245} {arXiv:hep-ph/0102245 [hep-ph]}
  \BibitemShut {NoStop}%
\bibitem [{\citenamefont {Dalley}\ and\ \citenamefont {van~de
  Sande}(2003)}]{Dalley:2002nj}%
  \BibitemOpen
  \bibfield  {author} {\bibinfo {author} {\bibfnamefont {S.}~\bibnamefont
  {Dalley}}\ and\ \bibinfo {author} {\bibfnamefont {B.}~\bibnamefont {van~de
  Sande}},\ }\href {\doibase 10.1103/PhysRevD.67.114507} {\bibfield  {journal}
  {\bibinfo  {journal} {Phys. Rev.}\ }\textbf {\bibinfo {volume} {D67}},\
  \bibinfo {pages} {114507} (\bibinfo {year} {2003})},\ \Eprint
  {http://arxiv.org/abs/hep-ph/0212086} {arXiv:hep-ph/0212086 [hep-ph]}
  \BibitemShut {NoStop}%
\bibitem [{\citenamefont {Burkardt}\ and\ \citenamefont
  {Dalley}(2002)}]{Burkardt:2001jg}%
  \BibitemOpen
  \bibfield  {author} {\bibinfo {author} {\bibfnamefont {M.}~\bibnamefont
  {Burkardt}}\ and\ \bibinfo {author} {\bibfnamefont {S.}~\bibnamefont
  {Dalley}},\ }\href {\doibase 10.1016/S0146-6410(02)00140-0} {\bibfield
  {journal} {\bibinfo  {journal} {Prog. Part. Nucl. Phys.}\ }\textbf {\bibinfo
  {volume} {48}},\ \bibinfo {pages} {317} (\bibinfo {year} {2002})},\ \Eprint
  {http://arxiv.org/abs/hep-ph/0112007} {arXiv:hep-ph/0112007 [hep-ph]}
  \BibitemShut {NoStop}%
\bibitem [{\citenamefont {Musch}\ \emph {et~al.}(2011)\citenamefont {Musch},
  \citenamefont {Hagler}, \citenamefont {Negele},\ and\ \citenamefont
  {Schafer}}]{Musch:2010ka}%
  \BibitemOpen
  \bibfield  {author} {\bibinfo {author} {\bibfnamefont {B.~U.}\ \bibnamefont
  {Musch}}, \bibinfo {author} {\bibfnamefont {P.}~\bibnamefont {Hagler}},
  \bibinfo {author} {\bibfnamefont {J.~W.}\ \bibnamefont {Negele}}, \ and\
  \bibinfo {author} {\bibfnamefont {A.}~\bibnamefont {Schafer}},\ }\href
  {\doibase 10.1103/PhysRevD.83.094507} {\bibfield  {journal} {\bibinfo
  {journal} {Phys. Rev.}\ }\textbf {\bibinfo {volume} {D83}},\ \bibinfo {pages}
  {094507} (\bibinfo {year} {2011})},\ \Eprint {http://arxiv.org/abs/1011.1213}
  {arXiv:1011.1213 [hep-lat]} \BibitemShut {NoStop}%
\bibitem [{\citenamefont {Ji}(2013)}]{Ji:2013dva}%
  \BibitemOpen
  \bibfield  {author} {\bibinfo {author} {\bibfnamefont {X.}~\bibnamefont
  {Ji}},\ }\href {\doibase 10.1103/PhysRevLett.110.262002} {\bibfield
  {journal} {\bibinfo  {journal} {Phys. Rev. Lett.}\ }\textbf {\bibinfo
  {volume} {110}},\ \bibinfo {pages} {262002} (\bibinfo {year} {2013})},\
  \Eprint {http://arxiv.org/abs/1305.1539} {arXiv:1305.1539 [hep-ph]}
  \BibitemShut {NoStop}%
\bibitem [{\citenamefont {Xiong}\ \emph {et~al.}(2014)\citenamefont {Xiong},
  \citenamefont {Ji}, \citenamefont {Zhang},\ and\ \citenamefont
  {Zhao}}]{Xiong:2013bka}%
  \BibitemOpen
  \bibfield  {author} {\bibinfo {author} {\bibfnamefont {X.}~\bibnamefont
  {Xiong}}, \bibinfo {author} {\bibfnamefont {X.}~\bibnamefont {Ji}}, \bibinfo
  {author} {\bibfnamefont {J.-H.}\ \bibnamefont {Zhang}}, \ and\ \bibinfo
  {author} {\bibfnamefont {Y.}~\bibnamefont {Zhao}},\ }\href {\doibase
  10.1103/PhysRevD.90.014051} {\bibfield  {journal} {\bibinfo  {journal} {Phys.
  Rev.}\ }\textbf {\bibinfo {volume} {D90}},\ \bibinfo {pages} {014051}
  (\bibinfo {year} {2014})},\ \Eprint {http://arxiv.org/abs/1310.7471}
  {arXiv:1310.7471 [hep-ph]} \BibitemShut {NoStop}%
\bibitem [{\citenamefont {Ji}(2014)}]{Ji:2014gla}%
  \BibitemOpen
  \bibfield  {author} {\bibinfo {author} {\bibfnamefont {X.}~\bibnamefont
  {Ji}},\ }\href {\doibase 10.1007/s11433-014-5492-3} {\bibfield  {journal}
  {\bibinfo  {journal} {Sci. China Phys. Mech. Astron.}\ }\textbf {\bibinfo
  {volume} {57}},\ \bibinfo {pages} {1407} (\bibinfo {year} {2014})},\ \Eprint
  {http://arxiv.org/abs/1404.6680} {arXiv:1404.6680 [hep-ph]} \BibitemShut
  {NoStop}%
\bibitem [{\citenamefont {Ma}\ and\ \citenamefont {Qiu}(2014)}]{Ma:2014jla}%
  \BibitemOpen
  \bibfield  {author} {\bibinfo {author} {\bibfnamefont {Y.-Q.}\ \bibnamefont
  {Ma}}\ and\ \bibinfo {author} {\bibfnamefont {J.-W.}\ \bibnamefont {Qiu}},\
  }\href@noop {} {\  (\bibinfo {year} {2014})},\ \Eprint
  {http://arxiv.org/abs/1404.6860} {arXiv:1404.6860 [hep-ph]} \BibitemShut
  {NoStop}%
\bibitem [{\citenamefont {Ji}\ and\ \citenamefont {Zhang}(2015)}]{Ji:2015jwa}%
  \BibitemOpen
  \bibfield  {author} {\bibinfo {author} {\bibfnamefont {X.}~\bibnamefont
  {Ji}}\ and\ \bibinfo {author} {\bibfnamefont {J.-H.}\ \bibnamefont {Zhang}},\
  }\href {\doibase 10.1103/PhysRevD.92.034006} {\bibfield  {journal} {\bibinfo
  {journal} {Phys. Rev.}\ }\textbf {\bibinfo {volume} {D92}},\ \bibinfo {pages}
  {034006} (\bibinfo {year} {2015})},\ \Eprint
  {http://arxiv.org/abs/1505.07699} {arXiv:1505.07699 [hep-ph]} \BibitemShut
  {NoStop}%
\bibitem [{\citenamefont {Ji}\ \emph {et~al.}(2015)\citenamefont {Ji},
  \citenamefont {Sch{\"a}fer}, \citenamefont {Xiong},\ and\ \citenamefont
  {Zhang}}]{Ji:2015qla}%
  \BibitemOpen
  \bibfield  {author} {\bibinfo {author} {\bibfnamefont {X.}~\bibnamefont
  {Ji}}, \bibinfo {author} {\bibfnamefont {A.}~\bibnamefont {Sch{\"a}fer}},
  \bibinfo {author} {\bibfnamefont {X.}~\bibnamefont {Xiong}}, \ and\ \bibinfo
  {author} {\bibfnamefont {J.-H.}\ \bibnamefont {Zhang}},\ }\href {\doibase
  10.1103/PhysRevD.92.014039} {\bibfield  {journal} {\bibinfo  {journal} {Phys.
  Rev.}\ }\textbf {\bibinfo {volume} {D92}},\ \bibinfo {pages} {014039}
  (\bibinfo {year} {2015})},\ \Eprint {http://arxiv.org/abs/1506.00248}
  {arXiv:1506.00248 [hep-ph]} \BibitemShut {NoStop}%
\bibitem [{\citenamefont
  {Radyushkin}(2017{\natexlab{a}})}]{Radyushkin:2016hsy}%
  \BibitemOpen
  \bibfield  {author} {\bibinfo {author} {\bibfnamefont {A.}~\bibnamefont
  {Radyushkin}},\ }\href {\doibase 10.1016/j.physletb.2017.02.019} {\bibfield
  {journal} {\bibinfo  {journal} {Phys. Lett.}\ }\textbf {\bibinfo {volume}
  {B767}},\ \bibinfo {pages} {314} (\bibinfo {year} {2017}{\natexlab{a}})},\
  \Eprint {http://arxiv.org/abs/1612.05170} {arXiv:1612.05170 [hep-ph]}
  \BibitemShut {NoStop}%
\bibitem [{\citenamefont {Monahan}\ and\ \citenamefont
  {Orginos}(2017)}]{Monahan:2016bvm}%
  \BibitemOpen
  \bibfield  {author} {\bibinfo {author} {\bibfnamefont {C.}~\bibnamefont
  {Monahan}}\ and\ \bibinfo {author} {\bibfnamefont {K.}~\bibnamefont
  {Orginos}},\ }\href {\doibase 10.1007/JHEP03(2017)116} {\bibfield  {journal}
  {\bibinfo  {journal} {JHEP}\ }\textbf {\bibinfo {volume} {03}},\ \bibinfo
  {pages} {116} (\bibinfo {year} {2017})},\ \Eprint
  {http://arxiv.org/abs/1612.01584} {arXiv:1612.01584 [hep-lat]} \BibitemShut
  {NoStop}%
\bibitem [{\citenamefont {Chen}\ \emph {et~al.}(2017)\citenamefont {Chen},
  \citenamefont {Ji},\ and\ \citenamefont {Zhang}}]{Chen:2016fxx}%
  \BibitemOpen
  \bibfield  {author} {\bibinfo {author} {\bibfnamefont {J.-W.}\ \bibnamefont
  {Chen}}, \bibinfo {author} {\bibfnamefont {X.}~\bibnamefont {Ji}}, \ and\
  \bibinfo {author} {\bibfnamefont {J.-H.}\ \bibnamefont {Zhang}},\ }\href
  {\doibase 10.1016/j.nuclphysb.2016.12.004} {\bibfield  {journal} {\bibinfo
  {journal} {Nucl. Phys.}\ }\textbf {\bibinfo {volume} {B915}},\ \bibinfo
  {pages} {1} (\bibinfo {year} {2017})},\ \Eprint
  {http://arxiv.org/abs/1609.08102} {arXiv:1609.08102 [hep-ph]} \BibitemShut
  {NoStop}%
\bibitem [{\citenamefont {Ji}\ \emph {et~al.}(2017)\citenamefont {Ji},
  \citenamefont {Zhang},\ and\ \citenamefont {Zhao}}]{Ji:2017rah}%
  \BibitemOpen
  \bibfield  {author} {\bibinfo {author} {\bibfnamefont {X.}~\bibnamefont
  {Ji}}, \bibinfo {author} {\bibfnamefont {J.-H.}\ \bibnamefont {Zhang}}, \
  and\ \bibinfo {author} {\bibfnamefont {Y.}~\bibnamefont {Zhao}},\ }\href
  {\doibase 10.1016/j.nuclphysb.2017.09.001} {\bibfield  {journal} {\bibinfo
  {journal} {Nucl. Phys.}\ }\textbf {\bibinfo {volume} {B924}},\ \bibinfo
  {pages} {366} (\bibinfo {year} {2017})},\ \Eprint
  {http://arxiv.org/abs/1706.07416} {arXiv:1706.07416 [hep-ph]} \BibitemShut
  {NoStop}%
\bibitem [{\citenamefont {Chen}\ \emph {et~al.}(2018)\citenamefont {Chen},
  \citenamefont {Ishikawa}, \citenamefont {Jin}, \citenamefont {Lin},
  \citenamefont {Yang}, \citenamefont {Zhang},\ and\ \citenamefont
  {Zhao}}]{Chen:2017mzz}%
  \BibitemOpen
  \bibfield  {author} {\bibinfo {author} {\bibfnamefont {J.-W.}\ \bibnamefont
  {Chen}}, \bibinfo {author} {\bibfnamefont {T.}~\bibnamefont {Ishikawa}},
  \bibinfo {author} {\bibfnamefont {L.}~\bibnamefont {Jin}}, \bibinfo {author}
  {\bibfnamefont {H.-W.}\ \bibnamefont {Lin}}, \bibinfo {author} {\bibfnamefont
  {Y.-B.}\ \bibnamefont {Yang}}, \bibinfo {author} {\bibfnamefont {J.-H.}\
  \bibnamefont {Zhang}}, \ and\ \bibinfo {author} {\bibfnamefont
  {Y.}~\bibnamefont {Zhao}},\ }\href {\doibase 10.1103/PhysRevD.97.014505}
  {\bibfield  {journal} {\bibinfo  {journal} {Phys. Rev.}\ }\textbf {\bibinfo
  {volume} {D97}},\ \bibinfo {pages} {014505} (\bibinfo {year} {2018})},\
  \Eprint {http://arxiv.org/abs/1706.01295} {arXiv:1706.01295 [hep-lat]}
  \BibitemShut {NoStop}%
\bibitem [{\citenamefont {Carlson}\ and\ \citenamefont
  {Freid}(2017)}]{Carlson:2017gpk}%
  \BibitemOpen
  \bibfield  {author} {\bibinfo {author} {\bibfnamefont {C.~E.}\ \bibnamefont
  {Carlson}}\ and\ \bibinfo {author} {\bibfnamefont {M.}~\bibnamefont
  {Freid}},\ }\href {\doibase 10.1103/PhysRevD.95.094504} {\bibfield  {journal}
  {\bibinfo  {journal} {Phys. Rev.}\ }\textbf {\bibinfo {volume} {D95}},\
  \bibinfo {pages} {094504} (\bibinfo {year} {2017})},\ \Eprint
  {http://arxiv.org/abs/1702.05775} {arXiv:1702.05775 [hep-ph]} \BibitemShut
  {NoStop}%
\bibitem [{\citenamefont {Briceño}\ \emph {et~al.}(2017)\citenamefont
  {Briceño}, \citenamefont {Hansen},\ and\ \citenamefont
  {Monahan}}]{Briceno:2017cpo}%
  \BibitemOpen
  \bibfield  {author} {\bibinfo {author} {\bibfnamefont {R.~A.}\ \bibnamefont
  {Briceño}}, \bibinfo {author} {\bibfnamefont {M.~T.}\ \bibnamefont
  {Hansen}}, \ and\ \bibinfo {author} {\bibfnamefont {C.~J.}\ \bibnamefont
  {Monahan}},\ }\href {\doibase 10.1103/PhysRevD.96.014502} {\bibfield
  {journal} {\bibinfo  {journal} {Phys. Rev.}\ }\textbf {\bibinfo {volume}
  {D96}},\ \bibinfo {pages} {014502} (\bibinfo {year} {2017})},\ \Eprint
  {http://arxiv.org/abs/1703.06072} {arXiv:1703.06072 [hep-lat]} \BibitemShut
  {NoStop}%
\bibitem [{\citenamefont {Rossi}\ and\ \citenamefont
  {Testa}(2017)}]{Rossi:2017muf}%
  \BibitemOpen
  \bibfield  {author} {\bibinfo {author} {\bibfnamefont {G.~C.}\ \bibnamefont
  {Rossi}}\ and\ \bibinfo {author} {\bibfnamefont {M.}~\bibnamefont {Testa}},\
  }\href {\doibase 10.1103/PhysRevD.96.014507} {\bibfield  {journal} {\bibinfo
  {journal} {Phys. Rev.}\ }\textbf {\bibinfo {volume} {D96}},\ \bibinfo {pages}
  {014507} (\bibinfo {year} {2017})},\ \Eprint
  {http://arxiv.org/abs/1706.04428} {arXiv:1706.04428 [hep-lat]} \BibitemShut
  {NoStop}%
\bibitem [{\citenamefont {Stewart}\ and\ \citenamefont
  {Zhao}(2017)}]{Stewart:2017tvs}%
  \BibitemOpen
  \bibfield  {author} {\bibinfo {author} {\bibfnamefont {I.~W.}\ \bibnamefont
  {Stewart}}\ and\ \bibinfo {author} {\bibfnamefont {Y.}~\bibnamefont {Zhao}},\
  }\href@noop {} {\  (\bibinfo {year} {2017})},\ \Eprint
  {http://arxiv.org/abs/1709.04933} {arXiv:1709.04933 [hep-ph]} \BibitemShut
  {NoStop}%
\bibitem [{\citenamefont
  {Radyushkin}(2017{\natexlab{b}})}]{Radyushkin:2017ffo}%
  \BibitemOpen
  \bibfield  {author} {\bibinfo {author} {\bibfnamefont {A.}~\bibnamefont
  {Radyushkin}},\ }\href {\doibase 10.1016/j.physletb.2017.05.024} {\bibfield
  {journal} {\bibinfo  {journal} {Phys. Lett.}\ }\textbf {\bibinfo {volume}
  {B770}},\ \bibinfo {pages} {514} (\bibinfo {year} {2017}{\natexlab{b}})},\
  \Eprint {http://arxiv.org/abs/1702.01726} {arXiv:1702.01726 [hep-ph]}
  \BibitemShut {NoStop}%
\bibitem [{\citenamefont {Alexandrou}\ \emph {et~al.}(2015)\citenamefont
  {Alexandrou}, \citenamefont {Cichy}, \citenamefont {Drach}, \citenamefont
  {Garcia-Ramos}, \citenamefont {Hadjiyiannakou}, \citenamefont {Jansen},
  \citenamefont {Steffens},\ and\ \citenamefont {Wiese}}]{Alexandrou:2015rja}%
  \BibitemOpen
  \bibfield  {author} {\bibinfo {author} {\bibfnamefont {C.}~\bibnamefont
  {Alexandrou}}, \bibinfo {author} {\bibfnamefont {K.}~\bibnamefont {Cichy}},
  \bibinfo {author} {\bibfnamefont {V.}~\bibnamefont {Drach}}, \bibinfo
  {author} {\bibfnamefont {E.}~\bibnamefont {Garcia-Ramos}}, \bibinfo {author}
  {\bibfnamefont {K.}~\bibnamefont {Hadjiyiannakou}}, \bibinfo {author}
  {\bibfnamefont {K.}~\bibnamefont {Jansen}}, \bibinfo {author} {\bibfnamefont
  {F.}~\bibnamefont {Steffens}}, \ and\ \bibinfo {author} {\bibfnamefont
  {C.}~\bibnamefont {Wiese}},\ }\href {\doibase 10.1103/PhysRevD.92.014502}
  {\bibfield  {journal} {\bibinfo  {journal} {Phys. Rev.}\ }\textbf {\bibinfo
  {volume} {D92}},\ \bibinfo {pages} {014502} (\bibinfo {year} {2015})},\
  \Eprint {http://arxiv.org/abs/1504.07455} {arXiv:1504.07455 [hep-lat]}
  \BibitemShut {NoStop}%
\bibitem [{\citenamefont {Alexandrou}\ \emph {et~al.}(2016)\citenamefont
  {Alexandrou}, \citenamefont {Cichy}, \citenamefont {Constantinou},
  \citenamefont {Hadjiyiannakou}, \citenamefont {Jansen}, \citenamefont
  {Steffens},\ and\ \citenamefont {Wiese}}]{Alexandrou:2016eyt}%
  \BibitemOpen
  \bibfield  {author} {\bibinfo {author} {\bibfnamefont {C.}~\bibnamefont
  {Alexandrou}}, \bibinfo {author} {\bibfnamefont {K.}~\bibnamefont {Cichy}},
  \bibinfo {author} {\bibfnamefont {M.}~\bibnamefont {Constantinou}}, \bibinfo
  {author} {\bibfnamefont {K.}~\bibnamefont {Hadjiyiannakou}}, \bibinfo
  {author} {\bibfnamefont {K.}~\bibnamefont {Jansen}}, \bibinfo {author}
  {\bibfnamefont {F.}~\bibnamefont {Steffens}}, \ and\ \bibinfo {author}
  {\bibfnamefont {C.}~\bibnamefont {Wiese}},\ }\bibfield  {booktitle} {\emph
  {\bibinfo {booktitle} {{Proceedings, 34th International Symposium on Lattice
  Field Theory (Lattice 2016): Southampton, UK, July 24-30, 2016}}},\
  }\href@noop {} {\bibfield  {journal} {\bibinfo  {journal} {PoS}\ }\textbf
  {\bibinfo {volume} {LATTICE2016}},\ \bibinfo {pages} {151} (\bibinfo {year}
  {2016})},\ \Eprint {http://arxiv.org/abs/1612.08728} {arXiv:1612.08728
  [hep-lat]} \BibitemShut {NoStop}%
\bibitem [{\citenamefont {Alexandrou}\ \emph
  {et~al.}(2017{\natexlab{a}})\citenamefont {Alexandrou}, \citenamefont
  {Cichy}, \citenamefont {Constantinou}, \citenamefont {Hadjiyiannakou},
  \citenamefont {Jansen}, \citenamefont {Steffens},\ and\ \citenamefont
  {Wiese}}]{Alexandrou:2016jqi}%
  \BibitemOpen
  \bibfield  {author} {\bibinfo {author} {\bibfnamefont {C.}~\bibnamefont
  {Alexandrou}}, \bibinfo {author} {\bibfnamefont {K.}~\bibnamefont {Cichy}},
  \bibinfo {author} {\bibfnamefont {M.}~\bibnamefont {Constantinou}}, \bibinfo
  {author} {\bibfnamefont {K.}~\bibnamefont {Hadjiyiannakou}}, \bibinfo
  {author} {\bibfnamefont {K.}~\bibnamefont {Jansen}}, \bibinfo {author}
  {\bibfnamefont {F.}~\bibnamefont {Steffens}}, \ and\ \bibinfo {author}
  {\bibfnamefont {C.}~\bibnamefont {Wiese}},\ }\href {\doibase
  10.1103/PhysRevD.96.014513} {\bibfield  {journal} {\bibinfo  {journal} {Phys.
  Rev.}\ }\textbf {\bibinfo {volume} {D96}},\ \bibinfo {pages} {014513}
  (\bibinfo {year} {2017}{\natexlab{a}})},\ \Eprint
  {http://arxiv.org/abs/1610.03689} {arXiv:1610.03689 [hep-lat]} \BibitemShut
  {NoStop}%
\bibitem [{\citenamefont {Alexandrou}\ \emph {et~al.}()\citenamefont
  {Alexandrou}, \citenamefont {Bacchio}, \citenamefont {Cichy}, \citenamefont
  {Constantinou}, \citenamefont {Hadjiyiannakou}, \citenamefont {Jansen},
  \citenamefont {Koutsou}, \citenamefont {Scapellato},\ and\ \citenamefont
  {Steffens}}]{Alexandrou:2017dzj}%
  \BibitemOpen
  \bibfield  {author} {\bibinfo {author} {\bibfnamefont {C.}~\bibnamefont
  {Alexandrou}}, \bibinfo {author} {\bibfnamefont {S.}~\bibnamefont {Bacchio}},
  \bibinfo {author} {\bibfnamefont {K.}~\bibnamefont {Cichy}}, \bibinfo
  {author} {\bibfnamefont {M.}~\bibnamefont {Constantinou}}, \bibinfo {author}
  {\bibfnamefont {K.}~\bibnamefont {Hadjiyiannakou}}, \bibinfo {author}
  {\bibfnamefont {K.}~\bibnamefont {Jansen}}, \bibinfo {author} {\bibfnamefont
  {G.}~\bibnamefont {Koutsou}}, \bibinfo {author} {\bibfnamefont
  {A.}~\bibnamefont {Scapellato}}, \ and\ \bibinfo {author} {\bibfnamefont
  {F.}~\bibnamefont {Steffens}},\ }in\ \href@noop {} {\emph {\bibinfo
  {booktitle} {{35th International Symposium on Lattice Field Theory (Lattice
  2017) Granada, Spain, June 18-24, 2017}}}}\BibitemShut {NoStop}%
\bibitem [{\citenamefont {Orginos}\ \emph {et~al.}(2017)\citenamefont
  {Orginos}, \citenamefont {Radyushkin}, \citenamefont {Karpie},\ and\
  \citenamefont {Zafeiropoulos}}]{Orginos:2017kos}%
  \BibitemOpen
  \bibfield  {author} {\bibinfo {author} {\bibfnamefont {K.}~\bibnamefont
  {Orginos}}, \bibinfo {author} {\bibfnamefont {A.}~\bibnamefont {Radyushkin}},
  \bibinfo {author} {\bibfnamefont {J.}~\bibnamefont {Karpie}}, \ and\ \bibinfo
  {author} {\bibfnamefont {S.}~\bibnamefont {Zafeiropoulos}},\ }\href {\doibase
  10.1103/PhysRevD.96.094503} {\bibfield  {journal} {\bibinfo  {journal} {Phys.
  Rev.}\ }\textbf {\bibinfo {volume} {D96}},\ \bibinfo {pages} {094503}
  (\bibinfo {year} {2017})},\ \Eprint {http://arxiv.org/abs/1706.05373}
  {arXiv:1706.05373 [hep-ph]} \BibitemShut {NoStop}%
\bibitem [{\citenamefont {Gamberg}\ \emph {et~al.}(2015)\citenamefont
  {Gamberg}, \citenamefont {Kang}, \citenamefont {Vitev},\ and\ \citenamefont
  {Xing}}]{Gamberg:2014zwa}%
  \BibitemOpen
  \bibfield  {author} {\bibinfo {author} {\bibfnamefont {L.}~\bibnamefont
  {Gamberg}}, \bibinfo {author} {\bibfnamefont {Z.-B.}\ \bibnamefont {Kang}},
  \bibinfo {author} {\bibfnamefont {I.}~\bibnamefont {Vitev}}, \ and\ \bibinfo
  {author} {\bibfnamefont {H.}~\bibnamefont {Xing}},\ }\href {\doibase
  10.1016/j.physletb.2015.02.021} {\bibfield  {journal} {\bibinfo  {journal}
  {Phys. Lett.}\ }\textbf {\bibinfo {volume} {B743}},\ \bibinfo {pages} {112}
  (\bibinfo {year} {2015})},\ \Eprint {http://arxiv.org/abs/1412.3401}
  {arXiv:1412.3401 [hep-ph]} \BibitemShut {NoStop}%
\bibitem [{\citenamefont {Miller}\ and\ \citenamefont
  {Tiburzi}(2010)}]{Miller:2009fc}%
  \BibitemOpen
  \bibfield  {author} {\bibinfo {author} {\bibfnamefont {G.~A.}\ \bibnamefont
  {Miller}}\ and\ \bibinfo {author} {\bibfnamefont {B.~C.}\ \bibnamefont
  {Tiburzi}},\ }\href {\doibase 10.1103/PhysRevC.81.035201} {\bibfield
  {journal} {\bibinfo  {journal} {Phys. Rev.}\ }\textbf {\bibinfo {volume}
  {C81}},\ \bibinfo {pages} {035201} (\bibinfo {year} {2010})},\ \Eprint
  {http://arxiv.org/abs/0911.3691} {arXiv:0911.3691 [nucl-th]} \BibitemShut
  {NoStop}%
\bibitem [{\citenamefont {Broniowski}\ \emph {et~al.}(2010)\citenamefont
  {Broniowski}, \citenamefont {Prelovsek}, \citenamefont {Santelj},\ and\
  \citenamefont {Ruiz~Arriola}}]{Broniowski:2009dt}%
  \BibitemOpen
  \bibfield  {author} {\bibinfo {author} {\bibfnamefont {W.}~\bibnamefont
  {Broniowski}}, \bibinfo {author} {\bibfnamefont {S.}~\bibnamefont
  {Prelovsek}}, \bibinfo {author} {\bibfnamefont {L.}~\bibnamefont {Santelj}},
  \ and\ \bibinfo {author} {\bibfnamefont {E.}~\bibnamefont {Ruiz~Arriola}},\
  }\href {\doibase 10.1016/j.physletb.2010.02.074} {\bibfield  {journal}
  {\bibinfo  {journal} {Phys. Lett.}\ }\textbf {\bibinfo {volume} {B686}},\
  \bibinfo {pages} {313} (\bibinfo {year} {2010})},\ \Eprint
  {http://arxiv.org/abs/0911.4705} {arXiv:0911.4705 [hep-ph]} \BibitemShut
  {NoStop}%
\bibitem [{\citenamefont {Ruiz~Arriola}\ and\ \citenamefont
  {Broniowski}(2010)}]{Arriola:2010up}%
  \BibitemOpen
  \bibfield  {author} {\bibinfo {author} {\bibfnamefont {E.}~\bibnamefont
  {Ruiz~Arriola}}\ and\ \bibinfo {author} {\bibfnamefont {W.}~\bibnamefont
  {Broniowski}},\ }\bibfield  {booktitle} {\emph {\bibinfo {booktitle}
  {{Proceedings, International Workshop on Relativistic hadronic and particle
  physics (Light Cone 2010): Valencia, Spain, June 14-18, 2010}}},\ }\href@noop
  {} {\bibfield  {journal} {\bibinfo  {journal} {PoS}\ }\textbf {\bibinfo
  {volume} {LC2010}},\ \bibinfo {pages} {041} (\bibinfo {year} {2010})},\
  \Eprint {http://arxiv.org/abs/1009.5781} {arXiv:1009.5781 [hep-ph]}
  \BibitemShut {NoStop}%
\bibitem [{\citenamefont {Miller}(2010)}]{Miller:2010nz}%
  \BibitemOpen
  \bibfield  {author} {\bibinfo {author} {\bibfnamefont {G.~A.}\ \bibnamefont
  {Miller}},\ }\href {\doibase 10.1146/annurev.nucl.012809.104508} {\bibfield
  {journal} {\bibinfo  {journal} {Ann. Rev. Nucl. Part. Sci.}\ }\textbf
  {\bibinfo {volume} {60}},\ \bibinfo {pages} {1} (\bibinfo {year} {2010})},\
  \Eprint {http://arxiv.org/abs/1002.0355} {arXiv:1002.0355 [nucl-th]}
  \BibitemShut {NoStop}%
\bibitem [{\citenamefont
  {Radyushkin}(2017{\natexlab{c}})}]{Radyushkin:2017cyf}%
  \BibitemOpen
  \bibfield  {author} {\bibinfo {author} {\bibfnamefont {A.~V.}\ \bibnamefont
  {Radyushkin}},\ }\href {\doibase 10.1103/PhysRevD.96.034025} {\bibfield
  {journal} {\bibinfo  {journal} {Phys. Rev.}\ }\textbf {\bibinfo {volume}
  {D96}},\ \bibinfo {pages} {034025} (\bibinfo {year} {2017}{\natexlab{c}})},\
  \Eprint {http://arxiv.org/abs/1705.01488} {arXiv:1705.01488 [hep-ph]}
  \BibitemShut {NoStop}%
\bibitem [{\citenamefont
  {Radyushkin}(2017{\natexlab{d}})}]{Radyushkin:2017gjd}%
  \BibitemOpen
  \bibfield  {author} {\bibinfo {author} {\bibfnamefont {A.~V.}\ \bibnamefont
  {Radyushkin}},\ }\href {\doibase 10.1103/PhysRevD.95.056020} {\bibfield
  {journal} {\bibinfo  {journal} {Phys. Rev.}\ }\textbf {\bibinfo {volume}
  {D95}},\ \bibinfo {pages} {056020} (\bibinfo {year} {2017}{\natexlab{d}})},\
  \Eprint {http://arxiv.org/abs/1701.02688} {arXiv:1701.02688 [hep-ph]}
  \BibitemShut {NoStop}%
\bibitem [{\citenamefont
  {Radyushkin}(2017{\natexlab{e}})}]{Radyushkin:2017lvu}%
  \BibitemOpen
  \bibfield  {author} {\bibinfo {author} {\bibfnamefont {A.~V.}\ \bibnamefont
  {Radyushkin}},\ }\href@noop {} {\  (\bibinfo {year} {2017}{\natexlab{e}})},\
  \Eprint {http://arxiv.org/abs/1710.08813} {arXiv:1710.08813 [hep-ph]}
  \BibitemShut {NoStop}%
\bibitem [{\citenamefont {Angeles-Martinez}\ \emph {et~al.}(2015)\citenamefont
  {Angeles-Martinez} \emph {et~al.}}]{Angeles-Martinez:2015sea}%
  \BibitemOpen
  \bibfield  {author} {\bibinfo {author} {\bibfnamefont {R.}~\bibnamefont
  {Angeles-Martinez}} \emph {et~al.},\ }\href {\doibase
  10.5506/APhysPolB.46.2501} {\bibfield  {journal} {\bibinfo  {journal} {Acta
  Phys. Polon.}\ }\textbf {\bibinfo {volume} {B46}},\ \bibinfo {pages} {2501}
  (\bibinfo {year} {2015})},\ \Eprint {http://arxiv.org/abs/1507.05267}
  {arXiv:1507.05267 [hep-ph]} \BibitemShut {NoStop}%
\bibitem [{\citenamefont {Ioffe}(1969)}]{Ioffe:1969kf}%
  \BibitemOpen
  \bibfield  {author} {\bibinfo {author} {\bibfnamefont {B.~L.}\ \bibnamefont
  {Ioffe}},\ }\href {\doibase 10.1016/0370-2693(69)90415-8} {\bibfield
  {journal} {\bibinfo  {journal} {Phys. Lett.}\ }\textbf {\bibinfo {volume}
  {30B}},\ \bibinfo {pages} {123} (\bibinfo {year} {1969})}\BibitemShut
  {NoStop}%
\bibitem [{\citenamefont {Braun}\ \emph {et~al.}(1995)\citenamefont {Braun},
  \citenamefont {Gornicki},\ and\ \citenamefont {Mankiewicz}}]{Braun:1994jq}%
  \BibitemOpen
  \bibfield  {author} {\bibinfo {author} {\bibfnamefont {V.}~\bibnamefont
  {Braun}}, \bibinfo {author} {\bibfnamefont {P.}~\bibnamefont {Gornicki}}, \
  and\ \bibinfo {author} {\bibfnamefont {L.}~\bibnamefont {Mankiewicz}},\
  }\href {\doibase 10.1103/PhysRevD.51.6036} {\bibfield  {journal} {\bibinfo
  {journal} {Phys. Rev.}\ }\textbf {\bibinfo {volume} {D51}},\ \bibinfo {pages}
  {6036} (\bibinfo {year} {1995})},\ \Eprint
  {http://arxiv.org/abs/hep-ph/9410318} {arXiv:hep-ph/9410318 [hep-ph]}
  \BibitemShut {NoStop}%
\bibitem [{\citenamefont {Karpie}\ \emph {et~al.}(2017)\citenamefont {Karpie},
  \citenamefont {Orginos}, \citenamefont {Radyushkin},\ and\ \citenamefont
  {Zafeiropoulos}}]{Karpie:2017bzm}%
  \BibitemOpen
  \bibfield  {author} {\bibinfo {author} {\bibfnamefont {J.}~\bibnamefont
  {Karpie}}, \bibinfo {author} {\bibfnamefont {K.}~\bibnamefont {Orginos}},
  \bibinfo {author} {\bibfnamefont {A.}~\bibnamefont {Radyushkin}}, \ and\
  \bibinfo {author} {\bibfnamefont {S.}~\bibnamefont {Zafeiropoulos}},\
  }\href@noop {} {\  (\bibinfo {year} {2017})},\ \Eprint
  {http://arxiv.org/abs/1710.08288} {arXiv:1710.08288 [hep-lat]} \BibitemShut
  {NoStop}%
\bibitem [{\citenamefont {Monahan}\ and\ \citenamefont
  {Orginos}()}]{Monahan:2017oof}%
  \BibitemOpen
  \bibfield  {author} {\bibinfo {author} {\bibfnamefont {C.}~\bibnamefont
  {Monahan}}\ and\ \bibinfo {author} {\bibfnamefont {K.}~\bibnamefont
  {Orginos}},\ }in\ \href@noop {} {\emph {\bibinfo {booktitle} {{35th
  International Symposium on Lattice Field Theory (Lattice 2017) Granada,
  Spain, June 18-24, 2017}}}}\BibitemShut {NoStop}%
\bibitem [{\citenamefont {D'Alesio}\ and\ \citenamefont
  {Murgia}(2004)}]{DAlesio:2004eso}%
  \BibitemOpen
  \bibfield  {author} {\bibinfo {author} {\bibfnamefont {U.}~\bibnamefont
  {D'Alesio}}\ and\ \bibinfo {author} {\bibfnamefont {F.}~\bibnamefont
  {Murgia}},\ }\href {\doibase 10.1103/PhysRevD.70.074009} {\bibfield
  {journal} {\bibinfo  {journal} {Phys. Rev.}\ }\textbf {\bibinfo {volume}
  {D70}},\ \bibinfo {pages} {074009} (\bibinfo {year} {2004})},\ \Eprint
  {http://arxiv.org/abs/hep-ph/0408092} {arXiv:hep-ph/0408092 [hep-ph]}
  \BibitemShut {NoStop}%
\bibitem [{\citenamefont {Schweitzer}\ \emph {et~al.}(2010)\citenamefont
  {Schweitzer}, \citenamefont {Teckentrup},\ and\ \citenamefont
  {Metz}}]{Schweitzer:2010tt}%
  \BibitemOpen
  \bibfield  {author} {\bibinfo {author} {\bibfnamefont {P.}~\bibnamefont
  {Schweitzer}}, \bibinfo {author} {\bibfnamefont {T.}~\bibnamefont
  {Teckentrup}}, \ and\ \bibinfo {author} {\bibfnamefont {A.}~\bibnamefont
  {Metz}},\ }\href {\doibase 10.1103/PhysRevD.81.094019} {\bibfield  {journal}
  {\bibinfo  {journal} {Phys. Rev.}\ }\textbf {\bibinfo {volume} {D81}},\
  \bibinfo {pages} {094019} (\bibinfo {year} {2010})},\ \Eprint
  {http://arxiv.org/abs/1003.2190} {arXiv:1003.2190 [hep-ph]} \BibitemShut
  {NoStop}%
\bibitem [{\citenamefont {Ciafaloni}(1988)}]{Ciafaloni:1987ur}%
  \BibitemOpen
  \bibfield  {author} {\bibinfo {author} {\bibfnamefont {M.}~\bibnamefont
  {Ciafaloni}},\ }\href {\doibase 10.1016/0550-3213(88)90380-X} {\bibfield
  {journal} {\bibinfo  {journal} {Nucl. Phys.}\ }\textbf {\bibinfo {volume}
  {B296}},\ \bibinfo {pages} {49} (\bibinfo {year} {1988})}\BibitemShut
  {NoStop}%
\bibitem [{\citenamefont {Catani}\ \emph
  {et~al.}(1990{\natexlab{a}})\citenamefont {Catani}, \citenamefont {Fiorani},\
  and\ \citenamefont {Marchesini}}]{Catani:1989yc}%
  \BibitemOpen
  \bibfield  {author} {\bibinfo {author} {\bibfnamefont {S.}~\bibnamefont
  {Catani}}, \bibinfo {author} {\bibfnamefont {F.}~\bibnamefont {Fiorani}}, \
  and\ \bibinfo {author} {\bibfnamefont {G.}~\bibnamefont {Marchesini}},\
  }\href {\doibase 10.1016/0370-2693(90)91938-8} {\bibfield  {journal}
  {\bibinfo  {journal} {Phys. Lett.}\ }\textbf {\bibinfo {volume} {B234}},\
  \bibinfo {pages} {339} (\bibinfo {year} {1990}{\natexlab{a}})}\BibitemShut
  {NoStop}%
\bibitem [{\citenamefont {Catani}\ \emph
  {et~al.}(1990{\natexlab{b}})\citenamefont {Catani}, \citenamefont {Fiorani},\
  and\ \citenamefont {Marchesini}}]{Catani:1989sg}%
  \BibitemOpen
  \bibfield  {author} {\bibinfo {author} {\bibfnamefont {S.}~\bibnamefont
  {Catani}}, \bibinfo {author} {\bibfnamefont {F.}~\bibnamefont {Fiorani}}, \
  and\ \bibinfo {author} {\bibfnamefont {G.}~\bibnamefont {Marchesini}},\
  }\href {\doibase 10.1016/0550-3213(90)90342-B} {\bibfield  {journal}
  {\bibinfo  {journal} {Nucl. Phys.}\ }\textbf {\bibinfo {volume} {B336}},\
  \bibinfo {pages} {18} (\bibinfo {year} {1990}{\natexlab{b}})}\BibitemShut
  {NoStop}%
\bibitem [{\citenamefont {Kwiecinski}(2002)}]{Kwiecinski:2002bx}%
  \BibitemOpen
  \bibfield  {author} {\bibinfo {author} {\bibfnamefont {J.}~\bibnamefont
  {Kwiecinski}},\ }\href@noop {} {\bibfield  {journal} {\bibinfo  {journal}
  {Acta Phys. Polon.}\ }\textbf {\bibinfo {volume} {B33}},\ \bibinfo {pages}
  {1809} (\bibinfo {year} {2002})},\ \Eprint
  {http://arxiv.org/abs/hep-ph/0203172} {arXiv:hep-ph/0203172 [hep-ph]}
  \BibitemShut {NoStop}%
\bibitem [{\citenamefont {Gawron}\ and\ \citenamefont
  {Kwiecinski}(2003)}]{Gawron:2002kc}%
  \BibitemOpen
  \bibfield  {author} {\bibinfo {author} {\bibfnamefont {A.}~\bibnamefont
  {Gawron}}\ and\ \bibinfo {author} {\bibfnamefont {J.}~\bibnamefont
  {Kwiecinski}},\ }\href@noop {} {\bibfield  {journal} {\bibinfo  {journal}
  {Acta Phys. Polon.}\ }\textbf {\bibinfo {volume} {B34}},\ \bibinfo {pages}
  {133} (\bibinfo {year} {2003})},\ \Eprint
  {http://arxiv.org/abs/hep-ph/0207299} {arXiv:hep-ph/0207299 [hep-ph]}
  \BibitemShut {NoStop}%
\bibitem [{\citenamefont {Gawron}\ \emph {et~al.}(2003)\citenamefont {Gawron},
  \citenamefont {Kwiecinski},\ and\ \citenamefont
  {Broniowski}}]{Gawron:2003qg}%
  \BibitemOpen
  \bibfield  {author} {\bibinfo {author} {\bibfnamefont {A.}~\bibnamefont
  {Gawron}}, \bibinfo {author} {\bibfnamefont {J.}~\bibnamefont {Kwiecinski}},
  \ and\ \bibinfo {author} {\bibfnamefont {W.}~\bibnamefont {Broniowski}},\
  }\href {\doibase 10.1103/PhysRevD.68.054001} {\bibfield  {journal} {\bibinfo
  {journal} {Phys. Rev.}\ }\textbf {\bibinfo {volume} {D68}},\ \bibinfo {pages}
  {054001} (\bibinfo {year} {2003})},\ \Eprint
  {http://arxiv.org/abs/hep-ph/0305219} {arXiv:hep-ph/0305219 [hep-ph]}
  \BibitemShut {NoStop}%
\bibitem [{\citenamefont {Ruiz~Arriola}\ and\ \citenamefont
  {Broniowski}(2004)}]{RuizArriola:2004ui}%
  \BibitemOpen
  \bibfield  {author} {\bibinfo {author} {\bibfnamefont {E.}~\bibnamefont
  {Ruiz~Arriola}}\ and\ \bibinfo {author} {\bibfnamefont {W.}~\bibnamefont
  {Broniowski}},\ }\href {\doibase 10.1103/PhysRevD.70.034012} {\bibfield
  {journal} {\bibinfo  {journal} {Phys. Rev.}\ }\textbf {\bibinfo {volume}
  {D70}},\ \bibinfo {pages} {034012} (\bibinfo {year} {2004})},\ \Eprint
  {http://arxiv.org/abs/hep-ph/0404008} {arXiv:hep-ph/0404008 [hep-ph]}
  \BibitemShut {NoStop}%
\bibitem [{\citenamefont {Jaffe}(1983)}]{Jaffe:1983hp}%
  \BibitemOpen
  \bibfield  {author} {\bibinfo {author} {\bibfnamefont {R.~L.}\ \bibnamefont
  {Jaffe}},\ }\href {\doibase 10.1016/0550-3213(83)90361-9} {\bibfield
  {journal} {\bibinfo  {journal} {Nucl. Phys.}\ }\textbf {\bibinfo {volume}
  {B229}},\ \bibinfo {pages} {205} (\bibinfo {year} {1983})}\BibitemShut
  {NoStop}%
\bibitem [{\citenamefont {Jaffe}(1985)}]{Jaffe:1985je}%
  \BibitemOpen
  \bibfield  {author} {\bibinfo {author} {\bibfnamefont {R.~L.}\ \bibnamefont
  {Jaffe}},\ }in\ \href@noop {} {\emph {\bibinfo {booktitle} {{Proceedings,
  Research Program at CEBAF I: Report of the 1985 Summer Study Group, June 10 -
  August 30, 1985}}}}\ (\bibinfo {year} {1985})\BibitemShut {NoStop}%
\bibitem [{\citenamefont {Ji}(1998)}]{Ji:1998pc}%
  \BibitemOpen
  \bibfield  {author} {\bibinfo {author} {\bibfnamefont {X.-D.}\ \bibnamefont
  {Ji}},\ }\href {\doibase 10.1088/0954-3899/24/7/002} {\bibfield  {journal}
  {\bibinfo  {journal} {J. Phys.}\ }\textbf {\bibinfo {volume} {G24}},\
  \bibinfo {pages} {1181} (\bibinfo {year} {1998})},\ \Eprint
  {http://arxiv.org/abs/hep-ph/9807358} {arXiv:hep-ph/9807358 [hep-ph]}
  \BibitemShut {NoStop}%
\bibitem [{\citenamefont {Boer}\ \emph {et~al.}(2011)\citenamefont {Boer},
  \citenamefont {Gamberg}, \citenamefont {Musch},\ and\ \citenamefont
  {Prokudin}}]{Boer:2011xd}%
  \BibitemOpen
  \bibfield  {author} {\bibinfo {author} {\bibfnamefont {D.}~\bibnamefont
  {Boer}}, \bibinfo {author} {\bibfnamefont {L.}~\bibnamefont {Gamberg}},
  \bibinfo {author} {\bibfnamefont {B.}~\bibnamefont {Musch}}, \ and\ \bibinfo
  {author} {\bibfnamefont {A.}~\bibnamefont {Prokudin}},\ }\href {\doibase
  10.1007/JHEP10(2011)021} {\bibfield  {journal} {\bibinfo  {journal} {JHEP}\
  }\textbf {\bibinfo {volume} {10}},\ \bibinfo {pages} {021} (\bibinfo {year}
  {2011})},\ \Eprint {http://arxiv.org/abs/1107.5294} {arXiv:1107.5294
  [hep-ph]} \BibitemShut {NoStop}%
\bibitem [{\citenamefont {Bacchetta}\ \emph {et~al.}(2008)\citenamefont
  {Bacchetta}, \citenamefont {Conti},\ and\ \citenamefont
  {Radici}}]{Bacchetta:2008af}%
  \BibitemOpen
  \bibfield  {author} {\bibinfo {author} {\bibfnamefont {A.}~\bibnamefont
  {Bacchetta}}, \bibinfo {author} {\bibfnamefont {F.}~\bibnamefont {Conti}}, \
  and\ \bibinfo {author} {\bibfnamefont {M.}~\bibnamefont {Radici}},\ }\href
  {\doibase 10.1103/PhysRevD.78.074010} {\bibfield  {journal} {\bibinfo
  {journal} {Phys. Rev.}\ }\textbf {\bibinfo {volume} {D78}},\ \bibinfo {pages}
  {074010} (\bibinfo {year} {2008})},\ \Eprint {http://arxiv.org/abs/0807.0323}
  {arXiv:0807.0323 [hep-ph]} \BibitemShut {NoStop}%
\bibitem [{\citenamefont {Wakamatsu}(2009)}]{Wakamatsu:2009fn}%
  \BibitemOpen
  \bibfield  {author} {\bibinfo {author} {\bibfnamefont {M.}~\bibnamefont
  {Wakamatsu}},\ }\href {\doibase 10.1103/PhysRevD.79.094028} {\bibfield
  {journal} {\bibinfo  {journal} {Phys. Rev.}\ }\textbf {\bibinfo {volume}
  {D79}},\ \bibinfo {pages} {094028} (\bibinfo {year} {2009})},\ \Eprint
  {http://arxiv.org/abs/0903.1886} {arXiv:0903.1886 [hep-ph]} \BibitemShut
  {NoStop}%
\bibitem [{\citenamefont {Weigel}\ \emph {et~al.}(1999)\citenamefont {Weigel},
  \citenamefont {Ruiz~Arriola},\ and\ \citenamefont {Gamberg}}]{Weigel:1999pc}%
  \BibitemOpen
  \bibfield  {author} {\bibinfo {author} {\bibfnamefont {H.}~\bibnamefont
  {Weigel}}, \bibinfo {author} {\bibfnamefont {E.}~\bibnamefont
  {Ruiz~Arriola}}, \ and\ \bibinfo {author} {\bibfnamefont {L.~P.}\
  \bibnamefont {Gamberg}},\ }\href@noop {} {\bibfield  {journal} {\bibinfo
  {journal} {Nucl. Phys.}\ }\textbf {\bibinfo {volume} {B560}},\ \bibinfo
  {pages} {383} (\bibinfo {year} {1999})},\ \Eprint
  {http://arxiv.org/abs/hep-ph/9905329} {hep-ph/9905329} \BibitemShut {NoStop}%
\bibitem [{\citenamefont {Ruiz~Arriola}\ and\ \citenamefont
  {Broniowski}(2003{\natexlab{a}})}]{RuizArriola:2003wi}%
  \BibitemOpen
  \bibfield  {author} {\bibinfo {author} {\bibfnamefont {E.}~\bibnamefont
  {Ruiz~Arriola}}\ and\ \bibinfo {author} {\bibfnamefont {W.}~\bibnamefont
  {Broniowski}},\ }in\ \href@noop {} {\emph {\bibinfo {booktitle} {{Light cone
  physics: Hadrons and beyond: Proceedings. 2003}}}}\ (\bibinfo {year} {2003})\
  \Eprint {http://arxiv.org/abs/hep-ph/0310044} {arXiv:hep-ph/0310044 [hep-ph]}
  \BibitemShut {NoStop}%
\bibitem [{\citenamefont {Ruiz~Arriola}\ and\ \citenamefont
  {Broniowski}(2003{\natexlab{b}})}]{RuizArriola:2003bs}%
  \BibitemOpen
  \bibfield  {author} {\bibinfo {author} {\bibfnamefont {E.}~\bibnamefont
  {Ruiz~Arriola}}\ and\ \bibinfo {author} {\bibfnamefont {W.}~\bibnamefont
  {Broniowski}},\ }\href {\doibase 10.1103/PhysRevD.67.074021} {\bibfield
  {journal} {\bibinfo  {journal} {Phys. Rev.}\ }\textbf {\bibinfo {volume}
  {D67}},\ \bibinfo {pages} {074021} (\bibinfo {year} {2003}{\natexlab{b}})},\
  \Eprint {http://arxiv.org/abs/hep-ph/0301202} {arXiv:hep-ph/0301202 [hep-ph]}
  \BibitemShut {NoStop}%
\bibitem [{\citenamefont {Noguera}\ and\ \citenamefont
  {Scopetta}(2015)}]{Noguera:2015iia}%
  \BibitemOpen
  \bibfield  {author} {\bibinfo {author} {\bibfnamefont {S.}~\bibnamefont
  {Noguera}}\ and\ \bibinfo {author} {\bibfnamefont {S.}~\bibnamefont
  {Scopetta}},\ }\href {\doibase 10.1007/JHEP11(2015)102} {\bibfield  {journal}
  {\bibinfo  {journal} {JHEP}\ }\textbf {\bibinfo {volume} {11}},\ \bibinfo
  {pages} {102} (\bibinfo {year} {2015})},\ \Eprint
  {http://arxiv.org/abs/1508.01061} {arXiv:1508.01061 [hep-ph]} \BibitemShut
  {NoStop}%
\bibitem [{\citenamefont {Melis}(2015)}]{Melis:2014pna}%
  \BibitemOpen
  \bibfield  {author} {\bibinfo {author} {\bibfnamefont {S.}~\bibnamefont
  {Melis}},\ }\bibfield  {booktitle} {\emph {\bibinfo {booktitle}
  {{Proceedings, 4th International Workshop on Transverse Polarization
  Phenomena in Hard Processes (Transversity 2014): Cagliari, Italy, June 9-13,
  2014}}},\ }\href {\doibase 10.1051/epjconf/20158501001} {\bibfield  {journal}
  {\bibinfo  {journal} {EPJ Web Conf.}\ }\textbf {\bibinfo {volume} {85}},\
  \bibinfo {pages} {01001} (\bibinfo {year} {2015})},\ \Eprint
  {http://arxiv.org/abs/1412.1719} {arXiv:1412.1719 [hep-ph]} \BibitemShut
  {NoStop}%
\bibitem [{\citenamefont {Bacchetta}\ \emph {et~al.}(2017)\citenamefont
  {Bacchetta}, \citenamefont {Delcarro}, \citenamefont {Pisano}, \citenamefont
  {Radici},\ and\ \citenamefont {Signori}}]{Bacchetta:2017gcc}%
  \BibitemOpen
  \bibfield  {author} {\bibinfo {author} {\bibfnamefont {A.}~\bibnamefont
  {Bacchetta}}, \bibinfo {author} {\bibfnamefont {F.}~\bibnamefont {Delcarro}},
  \bibinfo {author} {\bibfnamefont {C.}~\bibnamefont {Pisano}}, \bibinfo
  {author} {\bibfnamefont {M.}~\bibnamefont {Radici}}, \ and\ \bibinfo {author}
  {\bibfnamefont {A.}~\bibnamefont {Signori}},\ }\href {\doibase
  10.1007/JHEP06(2017)081} {\bibfield  {journal} {\bibinfo  {journal} {JHEP}\
  }\textbf {\bibinfo {volume} {06}},\ \bibinfo {pages} {081} (\bibinfo {year}
  {2017})},\ \Eprint {http://arxiv.org/abs/1703.10157} {arXiv:1703.10157
  [hep-ph]} \BibitemShut {NoStop}%
\bibitem [{\citenamefont {Ball}\ \emph {et~al.}(2013)\citenamefont {Ball} \emph
  {et~al.}}]{Ball:2012cx}%
  \BibitemOpen
  \bibfield  {author} {\bibinfo {author} {\bibfnamefont {R.~D.}\ \bibnamefont
  {Ball}} \emph {et~al.},\ }\href {\doibase 10.1016/j.nuclphysb.2012.10.003}
  {\bibfield  {journal} {\bibinfo  {journal} {Nucl. Phys.}\ }\textbf {\bibinfo
  {volume} {B867}},\ \bibinfo {pages} {244} (\bibinfo {year} {2013})},\ \Eprint
  {http://arxiv.org/abs/1207.1303} {arXiv:1207.1303 [hep-ph]} \BibitemShut
  {NoStop}%
\bibitem [{\citenamefont {Broniowski}\ \emph {et~al.}(2008)\citenamefont
  {Broniowski}, \citenamefont {Ruiz~Arriola},\ and\ \citenamefont
  {Golec-Biernat}}]{Broniowski:2007si}%
  \BibitemOpen
  \bibfield  {author} {\bibinfo {author} {\bibfnamefont {W.}~\bibnamefont
  {Broniowski}}, \bibinfo {author} {\bibfnamefont {E.}~\bibnamefont
  {Ruiz~Arriola}}, \ and\ \bibinfo {author} {\bibfnamefont {K.}~\bibnamefont
  {Golec-Biernat}},\ }\href {\doibase 10.1103/PhysRevD.77.034023} {\bibfield
  {journal} {\bibinfo  {journal} {Phys.Rev.}\ }\textbf {\bibinfo {volume}
  {D77}},\ \bibinfo {pages} {034023} (\bibinfo {year} {2008})},\ \Eprint
  {http://arxiv.org/abs/0712.1012} {arXiv:0712.1012 [hep-ph]} \BibitemShut
  {NoStop}%
\bibitem [{\citenamefont {Gribov}\ and\ \citenamefont
  {Lipatov}(1972)}]{Gribov:1972ri}%
  \BibitemOpen
  \bibfield  {author} {\bibinfo {author} {\bibfnamefont {V.~N.}\ \bibnamefont
  {Gribov}}\ and\ \bibinfo {author} {\bibfnamefont {L.~N.}\ \bibnamefont
  {Lipatov}},\ }\href@noop {} {\bibfield  {journal} {\bibinfo  {journal} {Sov.
  J. Nucl. Phys.}\ }\textbf {\bibinfo {volume} {15}},\ \bibinfo {pages} {438}
  (\bibinfo {year} {1972})},\ \bibinfo {note} {[Yad.
  Fiz.15,781(1972)]}\BibitemShut {NoStop}%
\bibitem [{\citenamefont {Dokshitzer}(1977)}]{Dokshitzer:1977sg}%
  \BibitemOpen
  \bibfield  {author} {\bibinfo {author} {\bibfnamefont {Y.~L.}\ \bibnamefont
  {Dokshitzer}},\ }\href@noop {} {\bibfield  {journal} {\bibinfo  {journal}
  {Sov. Phys. JETP}\ }\textbf {\bibinfo {volume} {46}},\ \bibinfo {pages} {641}
  (\bibinfo {year} {1977})},\ \bibinfo {note} {[Zh. Eksp. Teor.
  Fiz.73,1216(1977)]}\BibitemShut {NoStop}%
\bibitem [{\citenamefont {Altarelli}\ and\ \citenamefont
  {Parisi}(1977)}]{Altarelli:1977zs}%
  \BibitemOpen
  \bibfield  {author} {\bibinfo {author} {\bibfnamefont {G.}~\bibnamefont
  {Altarelli}}\ and\ \bibinfo {author} {\bibfnamefont {G.}~\bibnamefont
  {Parisi}},\ }\href {\doibase 10.1016/0550-3213(77)90384-4} {\bibfield
  {journal} {\bibinfo  {journal} {Nucl. Phys.}\ }\textbf {\bibinfo {volume}
  {B126}},\ \bibinfo {pages} {298} (\bibinfo {year} {1977})}\BibitemShut
  {NoStop}%
\bibitem [{\citenamefont {Lipatov}(1976)}]{Lipatov:1976zz}%
  \BibitemOpen
  \bibfield  {author} {\bibinfo {author} {\bibfnamefont {L.~N.}\ \bibnamefont
  {Lipatov}},\ }\href@noop {} {\bibfield  {journal} {\bibinfo  {journal} {Sov.
  J. Nucl. Phys.}\ }\textbf {\bibinfo {volume} {23}},\ \bibinfo {pages} {338}
  (\bibinfo {year} {1976})},\ \bibinfo {note} {[Yad.
  Fiz.23,642(1976)]}\BibitemShut {NoStop}%
\bibitem [{\citenamefont {Kuraev}\ \emph {et~al.}(1977)\citenamefont {Kuraev},
  \citenamefont {Lipatov},\ and\ \citenamefont {Fadin}}]{Kuraev:1977fs}%
  \BibitemOpen
  \bibfield  {author} {\bibinfo {author} {\bibfnamefont {E.~A.}\ \bibnamefont
  {Kuraev}}, \bibinfo {author} {\bibfnamefont {L.~N.}\ \bibnamefont {Lipatov}},
  \ and\ \bibinfo {author} {\bibfnamefont {V.~S.}\ \bibnamefont {Fadin}},\
  }\href@noop {} {\bibfield  {journal} {\bibinfo  {journal} {Sov. Phys. JETP}\
  }\textbf {\bibinfo {volume} {45}},\ \bibinfo {pages} {199} (\bibinfo {year}
  {1977})},\ \bibinfo {note} {[Zh. Eksp. Teor. Fiz.72,377(1977)]}\BibitemShut
  {NoStop}%
\bibitem [{\citenamefont {Balitsky}\ and\ \citenamefont
  {Lipatov}(1978)}]{Balitsky:1978ic}%
  \BibitemOpen
  \bibfield  {author} {\bibinfo {author} {\bibfnamefont {I.~I.}\ \bibnamefont
  {Balitsky}}\ and\ \bibinfo {author} {\bibfnamefont {L.~N.}\ \bibnamefont
  {Lipatov}},\ }\href@noop {} {\bibfield  {journal} {\bibinfo  {journal} {Sov.
  J. Nucl. Phys.}\ }\textbf {\bibinfo {volume} {28}},\ \bibinfo {pages} {822}
  (\bibinfo {year} {1978})},\ \bibinfo {note} {[Yad.
  Fiz.28,1597(1978)]}\BibitemShut {NoStop}%
\bibitem [{\citenamefont {Golec-Biernat}\ \emph {et~al.}(2007)\citenamefont
  {Golec-Biernat}, \citenamefont {Jadach}, \citenamefont {Placzek},
  \citenamefont {Stephens},\ and\ \citenamefont
  {Skrzypek}}]{GolecBiernat:2007pu}%
  \BibitemOpen
  \bibfield  {author} {\bibinfo {author} {\bibfnamefont {K.~J.}\ \bibnamefont
  {Golec-Biernat}}, \bibinfo {author} {\bibfnamefont {S.}~\bibnamefont
  {Jadach}}, \bibinfo {author} {\bibfnamefont {W.}~\bibnamefont {Placzek}},
  \bibinfo {author} {\bibfnamefont {P.}~\bibnamefont {Stephens}}, \ and\
  \bibinfo {author} {\bibfnamefont {M.}~\bibnamefont {Skrzypek}},\ }\href@noop
  {} {\bibfield  {journal} {\bibinfo  {journal} {Acta Phys. Polon.}\ }\textbf
  {\bibinfo {volume} {B38}},\ \bibinfo {pages} {3149} (\bibinfo {year}
  {2007})},\ \Eprint {http://arxiv.org/abs/hep-ph/0703317}
  {arXiv:hep-ph/0703317 [hep-ph]} \BibitemShut {NoStop}%
\bibitem [{\citenamefont {Gustafson}\ \emph {et~al.}(2002)\citenamefont
  {Gustafson}, \citenamefont {Lonnblad},\ and\ \citenamefont
  {Miu}}]{Gustafson:2002jy}%
  \BibitemOpen
  \bibfield  {author} {\bibinfo {author} {\bibfnamefont {G.}~\bibnamefont
  {Gustafson}}, \bibinfo {author} {\bibfnamefont {L.}~\bibnamefont {Lonnblad}},
  \ and\ \bibinfo {author} {\bibfnamefont {G.}~\bibnamefont {Miu}},\ }\href
  {\doibase 10.1088/1126-6708/2002/09/005} {\bibfield  {journal} {\bibinfo
  {journal} {JHEP}\ }\textbf {\bibinfo {volume} {09}},\ \bibinfo {pages} {005}
  (\bibinfo {year} {2002})},\ \Eprint {http://arxiv.org/abs/hep-ph/0206195}
  {arXiv:hep-ph/0206195 [hep-ph]} \BibitemShut {NoStop}%
\bibitem [{\citenamefont {Glück}\ \emph {et~al.}(1998)\citenamefont {Glück},
  \citenamefont {Reya},\ and\ \citenamefont {Vogt}}]{Gluck:1998xa}%
  \BibitemOpen
  \bibfield  {author} {\bibinfo {author} {\bibfnamefont {M.}~\bibnamefont
  {Glück}}, \bibinfo {author} {\bibfnamefont {E.}~\bibnamefont {Reya}}, \ and\
  \bibinfo {author} {\bibfnamefont {A.}~\bibnamefont {Vogt}},\ }\href {\doibase
  10.1007/s100529800978, 10.1007/s100520050289} {\bibfield  {journal} {\bibinfo
   {journal} {Eur. Phys. J.}\ }\textbf {\bibinfo {volume} {C5}},\ \bibinfo
  {pages} {461} (\bibinfo {year} {1998})},\ \Eprint
  {http://arxiv.org/abs/hep-ph/9806404} {arXiv:hep-ph/9806404 [hep-ph]}
  \BibitemShut {NoStop}%
\bibitem [{\citenamefont {Alexandrou}\ \emph
  {et~al.}(2017{\natexlab{b}})\citenamefont {Alexandrou}, \citenamefont
  {Cichy}, \citenamefont {Constantinou}, \citenamefont {Hadjiyiannakou},
  \citenamefont {Jansen}, \citenamefont {Panagopoulos},\ and\ \citenamefont
  {Steffens}}]{Alexandrou:2017huk}%
  \BibitemOpen
  \bibfield  {author} {\bibinfo {author} {\bibfnamefont {C.}~\bibnamefont
  {Alexandrou}}, \bibinfo {author} {\bibfnamefont {K.}~\bibnamefont {Cichy}},
  \bibinfo {author} {\bibfnamefont {M.}~\bibnamefont {Constantinou}}, \bibinfo
  {author} {\bibfnamefont {K.}~\bibnamefont {Hadjiyiannakou}}, \bibinfo
  {author} {\bibfnamefont {K.}~\bibnamefont {Jansen}}, \bibinfo {author}
  {\bibfnamefont {H.}~\bibnamefont {Panagopoulos}}, \ and\ \bibinfo {author}
  {\bibfnamefont {F.}~\bibnamefont {Steffens}},\ }\href {\doibase
  10.1016/j.nuclphysb.2017.08.012} {\bibfield  {journal} {\bibinfo  {journal}
  {Nucl. Phys.}\ }\textbf {\bibinfo {volume} {B923}},\ \bibinfo {pages} {394}
  (\bibinfo {year} {2017}{\natexlab{b}})},\ \Eprint
  {http://arxiv.org/abs/1706.00265} {arXiv:1706.00265 [hep-lat]} \BibitemShut
  {NoStop}%
\bibitem [{\citenamefont {Zhang}\ \emph {et~al.}(2017)\citenamefont {Zhang},
  \citenamefont {Chen}, \citenamefont {Ji}, \citenamefont {Jin},\ and\
  \citenamefont {Lin}}]{Zhang:2017bzy}%
  \BibitemOpen
  \bibfield  {author} {\bibinfo {author} {\bibfnamefont {J.-H.}\ \bibnamefont
  {Zhang}}, \bibinfo {author} {\bibfnamefont {J.-W.}\ \bibnamefont {Chen}},
  \bibinfo {author} {\bibfnamefont {X.}~\bibnamefont {Ji}}, \bibinfo {author}
  {\bibfnamefont {L.}~\bibnamefont {Jin}}, \ and\ \bibinfo {author}
  {\bibfnamefont {H.-W.}\ \bibnamefont {Lin}},\ }\href {\doibase
  10.1103/PhysRevD.95.094514} {\bibfield  {journal} {\bibinfo  {journal} {Phys.
  Rev.}\ }\textbf {\bibinfo {volume} {D95}},\ \bibinfo {pages} {094514}
  (\bibinfo {year} {2017})},\ \Eprint {http://arxiv.org/abs/1702.00008}
  {arXiv:1702.00008 [hep-lat]} \BibitemShut {NoStop}%
\bibitem [{\citenamefont {Broniowski}\ and\ \citenamefont
  {Ruiz~Arriola}(2017)}]{Broniowski:2017wbr}%
  \BibitemOpen
  \bibfield  {author} {\bibinfo {author} {\bibfnamefont {W.}~\bibnamefont
  {Broniowski}}\ and\ \bibinfo {author} {\bibfnamefont {E.}~\bibnamefont
  {Ruiz~Arriola}},\ }\href {\doibase 10.1016/j.physletb.2017.08.055} {\bibfield
   {journal} {\bibinfo  {journal} {Phys. Lett.}\ }\textbf {\bibinfo {volume}
  {B773}},\ \bibinfo {pages} {385} (\bibinfo {year} {2017})},\ \Eprint
  {http://arxiv.org/abs/1707.09588} {arXiv:1707.09588 [hep-ph]} \BibitemShut
  {NoStop}%
\bibitem [{\citenamefont {Sutton}\ \emph {et~al.}(1992)\citenamefont {Sutton},
  \citenamefont {Martin}, \citenamefont {Roberts},\ and\ \citenamefont
  {Stirling}}]{Sutton:1991ay}%
  \BibitemOpen
  \bibfield  {author} {\bibinfo {author} {\bibfnamefont {P.~J.}\ \bibnamefont
  {Sutton}}, \bibinfo {author} {\bibfnamefont {A.~D.}\ \bibnamefont {Martin}},
  \bibinfo {author} {\bibfnamefont {R.~G.}\ \bibnamefont {Roberts}}, \ and\
  \bibinfo {author} {\bibfnamefont {W.~J.}\ \bibnamefont {Stirling}},\
  }\href@noop {} {\bibfield  {journal} {\bibinfo  {journal} {Phys. Rev.}\
  }\textbf {\bibinfo {volume} {D45}},\ \bibinfo {pages} {2349} (\bibinfo {year}
  {1992})}\BibitemShut {NoStop}%
\bibitem [{\citenamefont {Gluck}\ \emph {et~al.}(1999)\citenamefont {Gluck},
  \citenamefont {Reya},\ and\ \citenamefont {Schienbein}}]{Gluck:1999xe}%
  \BibitemOpen
  \bibfield  {author} {\bibinfo {author} {\bibfnamefont {M.}~\bibnamefont
  {Gluck}}, \bibinfo {author} {\bibfnamefont {E.}~\bibnamefont {Reya}}, \ and\
  \bibinfo {author} {\bibfnamefont {I.}~\bibnamefont {Schienbein}},\
  }\href@noop {} {\bibfield  {journal} {\bibinfo  {journal} {Eur. Phys. J.}\
  }\textbf {\bibinfo {volume} {C10}},\ \bibinfo {pages} {313} (\bibinfo {year}
  {1999})},\ \Eprint {http://arxiv.org/abs/hep-ph/9903288} {hep-ph/9903288}
  \BibitemShut {NoStop}%
\bibitem [{\citenamefont {Ruiz~Arriola}(2002)}]{RuizArriola:2002wr}%
  \BibitemOpen
  \bibfield  {author} {\bibinfo {author} {\bibfnamefont {E.}~\bibnamefont
  {Ruiz~Arriola}},\ }\bibfield  {booktitle} {\emph {\bibinfo {booktitle} {{42nd
  Cracow School of Theoretical Physics: 42nd Course 2002: Flavor Dynamics
  Zakopane, Poland, May 31-June 9, 2002}}},\ }\href@noop {} {\bibfield
  {journal} {\bibinfo  {journal} {Acta Phys. Polon.}\ }\textbf {\bibinfo
  {volume} {B33}},\ \bibinfo {pages} {4443} (\bibinfo {year} {2002})},\ \Eprint
  {http://arxiv.org/abs/hep-ph/0210007} {arXiv:hep-ph/0210007 [hep-ph]}
  \BibitemShut {NoStop}%
\bibitem [{\citenamefont {Davidson}\ and\ \citenamefont
  {Ruiz~Arriola}(1995)}]{Davidson:1994uv}%
  \BibitemOpen
  \bibfield  {author} {\bibinfo {author} {\bibfnamefont {R.}~\bibnamefont
  {Davidson}}\ and\ \bibinfo {author} {\bibfnamefont {E.}~\bibnamefont
  {Ruiz~Arriola}},\ }\href {\doibase 10.1016/0370-2693(95)00091-X} {\bibfield
  {journal} {\bibinfo  {journal} {Phys.Lett.}\ }\textbf {\bibinfo {volume}
  {B348}},\ \bibinfo {pages} {163} (\bibinfo {year} {1995})}\BibitemShut
  {NoStop}%
\end{thebibliography}%

\end{document}